\documentclass{article}

\usepackage[ngerman]{babel}        

\usepackage{ucs}
\usepackage[applemac]{inputenc}
\usepackage{lmodern}        
\usepackage[T1]{fontenc}    
\usepackage{textcomp}         

\usepackage[onehalfspacing]{setspace} 

\usepackage{graphicx, color}
\usepackage{float}

\usepackage{sidecap}
\usepackage[labelfont=bf, font=small, textfont=it, format=plain]{caption}
\usepackage[fleqn]{amsmath}

\usepackage{rotating}
\usepackage{url}

\usepackage{bibgerm}
\makeatletter
	\newenvironment{indentation}[3]%
	{\par\setlength{\parindent}{#3}
	\setlength{\leftmargin}{#1}       \setlength{\rightmargin}{#2}%
	\advance\linewidth -\leftmargin       \advance\linewidth -\rightmargin%
	\advance\@totalleftmargin\leftmargin  \@setpar{{\@@par}}%
	\parshape 1\@totalleftmargin \linewidth\ignorespaces}{\par}%
\makeatother

\begin{document}

\title{Sind Computersimulationen geeigneter als SchŸlerexperimente, um das VerstŠndnis elektrischer Grundschaltungen zu verbessern?
Eine Untersuchung im Physikunterricht der Oberstufe am Gymnasium }

\author{Bjšrn S. Schmekel}


\date{1. September 2009}

\maketitle

\tableofcontents

\section{EinfŸhrung}
Das hier vorliegende Untersuchungsvorhaben basiert auf Erfahrungen, die an der University of Colorado mit dem Einsatz von Computersimulation zu Unterrichtszwecken gemacht worden sind \cite{PhysRevSTPER.1.010103}. Die an der gleichen UniversitŠt ansŠssige Physics Education Research Group "`PER@C"' \cite{PERatC}, welche die grš§te Arbeitsgruppe in den Vereinigten Staaten zur Erforschung der Physikdidaktik darstellt, ist mit der Aufgabe angetreten, aufgrund von ernŸchternden Klausurergebnissen in den AnfŠngervorlesungen die Lehre im Fachbereich Physik zu verbessern. Das angeschlossene Physics Education Technology project "`PhET"' \cite{phet} untersucht, wie Technologie gewinnbringend im physikalischen Unterricht eingesetzt werden kann. 
Der Einsatz des Computers geht hierbei wesentlich Ÿber seine Ÿbliche Verwendung als Hilfsmittel (z.B. bei der Messdatenerfassung, der Lšsung mathematischer Probleme oder zu einfachen PrŠsentationszwecken) hinaus. Vielmehr werden SchŸlerexperimente (bzw. Experimente\footnote{Im folgenden Text wird nicht zwischen Versuchen und Experimenten unterschieden.}, die von Studenten selbst durchgefŸhrt werden) ersetzt durch entsprechende Simulationen dieser Experimente am Computer. Hierbei stellt sich eine Reihe von offensichtlichen Fragen, die nicht einfach umgangen werden kšnnen: Verliert der Physikunterricht seine AuthentizitŠt? Ist der Bezug zur Wirklichkeit beim Versicht auf reale Experimente noch gegeben? Handelt es sich mšglicherweise um eine reine Sparma§nahme, um teure Anschaffungen und Reparaturen von GerŠten zu vermeiden? Auch in der deutschsprachigen Literatur gibt es Vorbehalte gegenŸber den Einsatz von Computern \cite{Stoll:2002ys}. Zumindest in der Mittelstufe solle darauf geachtet werden, dass fundamentale experimentelle Fertigkeiten nicht vernachlŠssigt werden. In der Oberstufe hingegen stelle sich der Einsatz weniger problematisch dar, da diese Fertigkeiten als bekannt vorausgesetzt werden kšnnen sollten \cite{Mikelskis:2006zr}.
Auf alle diese Fragen soll kurz eingegangen werden. Das Hauptaugenmerk liegt jedoch sowohl bei der amerikanischen, als auch in der hier vorliegenden Arbeit auf der Fragestellung, ob das physikalische VerstŠndnis durch den Einsatz von Computersimulationen besser gefšrdert wird, als dies durch normale SchŸlerexperimente der Fall wŠre. 
\section{Die Untersuchung}
\subsection{Ergebnisse aus frŸheren Studien}
Zu diesem Thema existieren bisher nur relativ wenige Studien. Zacharia und Anderson \cite{Zacharia:2003ys} haben in einer Studie festgestellt, dass Studenten grš§ere Fortschritte beim konzeptionellen VerstŠndnis gemacht haben, wenn sie sich mit Hilfe von Computerprogrammen auf Laborversuche vorbereitet haben, als mit LehrbŸchern und klassischen †bungsaufgaben. In einer an einer Grundschule durchgefŸhrten Untersuchung sind Triona und Klahr \cite{Triona:2003zr} zum Ergebnis gekommen, dass unter sonst gleichen Bedingungen Computersimulationen und Videos von Demonstrationsexperimenten genauso gewinnbringend eingesetzt werden kšnnen wie richtige Experimente. Linn und Mitarbeiter \cite{Linn:2000ly} \cite{Linn:2004ve} haben bestŠtigen kšnnen, dass, als Hilfsmittel eingesetzt (z. B. zur Messdatenerfassung, zu PrŠsentationszwecken oder als Ersatz fŸr andere LaborgerŠte), Computer beim konzeptionellen VerstŠndnis unterstŸtzend wirken.
Finkelstein und Kollegen \cite{PhysRevSTPER.1.010103} haben an der University of Colorado in Zusammenarbeit mit dem PhET und der PER@C die vorhandenen Untersuchungen erweitert um die Fragestellungen, ob echte Versuche všllig durch Computersimulationen auf der Stufe eines undergraduates ersetzt werden kšnnen, ob dabei die gleichen Konzepte erlernt werden und ob dabei eventuell FŠhigkeiten auf der Strecke bleiben, die bei realen Experimenten erworben werden. Speziell geht es hierbei um die Fragestellung, ob derart betreute SchŸler\footnote{Im folgenden Text sind alle Bezeichnungen geschlechtsneutral zu verstehen. Mit ãSchŸlerÒ sind selbstverstŠndlich auch alle SchŸlerinnen gemeint.} / Studenten noch in der Lage sind, selber reale Versuche durchzufŸhren.
In den Vorlesungen sind einfachste Gleichstromschaltungen behandelt worden. Betrachtet worden sind drei Gruppen von Studenten, die entweder passende Experimente, Computersimulationen derartiger Experimente oder weder Experimente noch Computersimulationen durchgefŸhrt haben. In den Abschlussarbeiten haben die Studenten, die sich eingehend mit Simulationen beschŠftigt haben, am besten abgeschnitten. Die verwendeten Simulationsprogramme \cite{phet} sind sehr visuell und bilden die physikalischen Prozesse entsprechend gŠngiger Modellvorstellungen ab. So wird z. B. der Stromfluss in den Zuleitungen durch animierte blaue Kugeln dargestellt. Zur Messung von Spannungen und Stršmen steht ein virtuelles Multimeter bereit, dessen PrŸfspitzen mit der Maus an die entsprechenden Knotenpunkte angesetzt werden kann. Das virtuelle Multimeter ist eine Abbildung eines echten Multimeters, welches die mit den PrŸfspitzen abgegriffene Spannung bzw. den Strom anzeigt. 
\begin{figure}
\begin{indentation}{70pt}{0pt}{0pt}
\includegraphics[width=256pt]{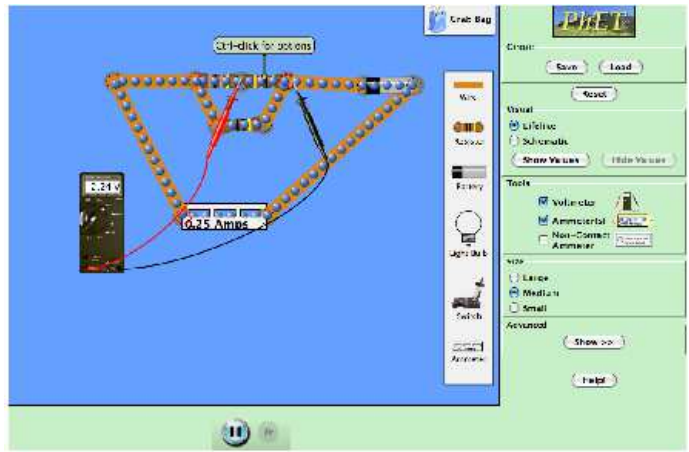}
\end{indentation}
\caption{Das Circuit Construction Lab}
\end{figure}

Aufgrund der ansprechenden Optik des Programms wird es auch in Vorlesungen zur Visualisierung eingesetzt. In den †bungsgruppen kann es dazu verwendet werden, von den Studenten geŠu§erte Hypothesen sofort zu ŸberprŸfen. Denkbar wŠre auch eine Verwendung des Programms bei der Bearbeitung der Hausaufgaben.
Beim Design der Simulationen ist darauf geachtet worden, dass sie das Entdecken von physikalischen ZusammenhŠngen fšrdern und nicht das blo§e BestŠtigen von vorher behandelten GesetzmŠ§igkeiten \cite{Aufschnaiter:2008sh} \cite{Thornton:1990sh} \cite{Laws:2004xd} \cite{McDermott:1996wb} \cite{Lippman:2003ys}.
Sie laden die Studenten zum eigenstŠndigen Experimentieren ein, z. B. durch die VerŠnderung der Betriebsparameter. Gleichzeitig ist der Gestaltungsspielraum jedoch stark eingeschrŠnkt. So ist es zum Beispiel nicht mšglich, Kabel verschiedener Farbe zu verwenden oder zum Beispiel eine VerstŠrkerschaltung zu bauen. Dadurch erfolgt eine Fokussierung auf das Wesentliche: Die Studenten sollen und kšnnen Variationen an der Schaltung vornehmen. Die Variationsmšglichkeiten sind jedoch eingeschrŠnkt, so dass diese tatsŠchlich zu einer tiefen Auseinandersetzung mit der Fragestellung fŸhren anstatt von dieser abzulenken. Die virtuellen Bauteile und MessgerŠte sind stark idealisiert, wodurch die Aufmerksamkeit auf ein konzeptionelles VerstŠndnis gelenkt wird. 
Ob sich die Ergebnisse hiermit auch allein erklŠren lassen oder ob vielleicht die Visualisierung die entscheidende Rolle spielt, geht aus der Studie nicht hervor. Erfahrungen mit den von der PhET entwickelten Simulationen legen die Vermutung nahe, dass die starke Visualisierung einen wesentlichen Einfluss auf den beobachteten positiven Effekt hat \cite{perkins:18} \cite{wieman2006}.

\subsection{Relevanz fŸr das deutsche Schulsystem}
Um die Schlussfolgerungen auf deutsche Schulen Ÿbertragen zu kšnnen, muss zuerst auf die Unterschiede zwischen dem deutschen und amerikanischen Bildungssystem eingegangen werden. Nach preschool (optional), kindergarten und elementary school erfolgt der Besuch der middle school bzw. junior high school und danach der (senior) high school, die mit der 12. Klasse endet. Die QualitŠt und die Lerninhalte dieser Schulen sind sehr unterschiedlich und hŠngen stark vom Bundesstaat, vom Schulbezirk und von den von den SchŸlern gewŠhlten Kursen ab. So sind die sogenannten advanced placement Kurse (AP-classes), deren Belegung freiwillig und deren Klausuren standardisiert sind, vom Inhalt her mit einem Grundkurs vergleichbar. Sie werden hŠufig nur in besseren Schulbezirken angeboten. Bei einem erfolgreichen Abschneiden in einem dieser Kurse kann in vielen FŠllen die erbrachte Leistung von einem college anerkannt werden. Der entsprechende Kurs muss dann im college nicht mehr belegt werden. Der Besuch eines weiterfŸhrenden colleges ist optional; er wird aufgrund der Arbeitsmarktsituation jedoch immer beliebter. In einem four-year college endet der erfolgreiche Besuch mit dem Erwerb eines bachelor degrees. Sofern dieser an einem guten college erworben wird, ist dieser Abschluss mit dem deutschen Vordiplom vergleichbar. Inhaltlich kann daher auch auf einem guten college erwartet werden, dass es sich bei den ersten zwei Jahren um eine Art gymnasiale Oberstufe handelt. 
Da selbst die Belegung von non-AP-physics in der Regel nicht verpflichtend ist, mŸssen colleges einfŸhrende Physikvorlesungen auf Mittelstufenniveau anbieten. Daher sind die Erkenntnisse der PER@C und des PhET fŸr deutsche Gymnasien durchaus relevant. 

\subsection{Ziele der Untersuchung}
In der hier vorliegenden Arbeit soll untersucht werden, ob eine EinschrŠnkung des Gestaltungsspielraumes bei der Entwicklung und Analyse von elektrischen Grundschaltungen und die damit verbundene Fokussierung auf die Kernaspekte allein bereits ausreicht, um eine Verbesserung des VerstŠndnisses elektrischer Grundschaltungen zu erzielen. Hierzu wird ein Simulationsprogramm eingesetzt, welches auch in der Industrie verwendet wird (SPICE). Da es nicht speziell fŸr Unterrichtszwecke entwickelt worden ist, fehlen Visualisierungen der zugrunde liegenden Modellvorstellung vollstŠndig. Dies ist beabsichtigt, um deren Einfluss im Vergleich zu frŸheren Arbeiten besser zu verstehen. Sofern deren Fehlen sich als zu nachteilig herausstellen sollte, kšnnen die Modellvorstellungen im UnterrichtsgesprŠch behandelt werden. Es wird gehofft, dass durch die Verwendung "`echter Software"' aus der Industrie die ArbeitsauftrŠge von den SchŸlern als authentisch eingestuft werden. 
Obwohl das Programm sehr umfangreich ist, kann der Gestaltungsspielraum fŸr eigene Experimente leicht eingeschrŠnkt werden. Hierzu wird auf die grafische MenŸfŸhrung verzichtet, die einen einfachen Zugang zu Hunderten von nichtbenštigten Funktionen ermšglicht und die SchŸler nur verwirren wŸrde. Stattdessen erfolgt die Bedienung wie zu Anfangszeiten der elektronischen Datenverarbeitung Ÿber Befehle, die in eine Textdatei geschrieben werden. Dadurch ist es mšglich zu steuern, mit welchen Befehlen die SchŸler experimentieren. 
Ziel der Untersuchung ist es zu bestimmen, ob SchŸlerexperimente durch entsprechende Simulationen am Computer ersetzt werden kšnnen und ob sich hieraus eventuell Vorteile fŸr das VerstŠndnis elektrischer Grundschaltungen ergeben. Auf eventuelle EinschrŠnkungen der Tauglichkeit wird geachtet. Durch die Eliminierung des Faktors der grafischen Visualisierung, welcher den SchŸlern jederzeit eine geeignete Modellbildung zur VerfŸgung stellt und welcher bei einer frŸheren Untersuchung als nicht unwesentlich angesehen wird, soll au§erdem versucht werden, die Frage zu klŠren, worin eventuelle Verbesserungen gegenŸber traditionellen SchŸlerexperimenten begrŸndet liegen.

\subsection{Hypothesen}
Aufgrund frŸherer Untersuchungsergebnisse werden fŸr den Einsatz von Computersimulationen anstelle von Experimenten die folgenden Hypothesen aufgestellt:

SchŸler erwerben ein tieferes VerstŠndnis fŸr elektrische Grundschaltungen bestehend aus WiderstŠnden und Kondensatoren und kšnnen Routineaufgaben zu diesem Themenkomplex besser bearbeiten.
\begin{itemize}
\item SchŸler kšnnen Schaltungen eigenstŠndig planen und diese mit Hilfe eines Simulationsprogramms untersuchen.
\item SchŸler kšnnen sich quantitativ intensiver mit der Materie auseinander setzen indem Routinearbeit reduziert und dadurch die Motivation gesteigert wird.
\item SchŸler planen Simulationen, die mit den vorhandenen Materialien nicht als Experiment durchfŸhrbar wŠren.
\item SchŸler arbeiten konzentrierter, indem ihre Aufmerksamkeit auf die relevanten Fragestellungen fokussiert wird. 
\end{itemize}
\subsection{Datenquellen}
Die dem Ergebnis zugrunde liegenden Daten stammen aus folgenden Quellen:
\begin{itemize}
\item Versuchsprotokolle mit genauen Zeitangaben
\item Feedbackbšgen
\item Videoanalyse
\item Tests
\end{itemize}

\subsection{Indikatoren}
Die folgenden Indikatoren sprechen fŸr den Einsatz von Simulationen:
\begin{itemize}
\item SchŸler schneiden bei entsprechenden Tests besser ab, die sowohl bekannte als auch unbekannte Probleme enthalten (aus Tests)
\item SchŸler brauchen weniger Zeit pro Experiment (aus Protokoll)
\item SchŸler machen mehr Experimente (leichte Variationen eingeschlossen) (aus Protokoll)
\item SchŸler bewerten den Einsatz von Simulationen positiv (aus Feedback)
\item SchŸler arbeiten zielgerichteter (keine Spielchen mit Kabel etc.) (aus Videoanalyse)
\end{itemize}
Da es darum geht, den SchŸlern Routinearbeit abzunehmen, sind der zweite und dritte Punkt wichtig um zu verstehen, ob tatsŠchlich in der gleichen Zeit mehr Aufgaben erledigt werden. Nur weil ein Versuch weniger Zeit in Anspruch nimmt, bedeutet dies nicht automatisch, dass die nun freigewordene Zeit auch produktiv genutzt wird. Daher mŸssen beide Indikatoren zusammen betrachtet werden. Ob die Methode tatsŠchlich ein zielgerichtetes Arbeiten fšrdert, kann teilweise schon aus der Betrachtung des Arbeitsverhaltens der SchŸler beantwortet werden. WŸrden z.B. die Computer in der nun verbleibenden Zeit nur zum Spielen der vorinstallierten Spiele genutzt werden, so kšnnte der Methode keine Effizienz zugesprochen werden. Auch Fragen zum Sozialverhalten der SchŸler lassen sich am einfachsten durch eine nochmalige Betrachtung der Unterrichtsstunde auf Video beantworten. Helfen sich die SchŸler gegenseitig? Wie sieht die Zusammenarbeit der SchŸler untereinander aus? Arbeiten auch SchŸler, die sich nicht fŸr das Fach begeistern kšnnen? Die fachlichen Kompetenzen lassen sich am einfachsten mit Hilfe von kurzen Tests ermitteln. Insbesondere ergibt sich so die Mšglichkeit, in einer standardisierten Art und Weise Unterschiede im Leistungsstand vor und nach der experimentellen Arbeit bzw. der Arbeit am Computer feststellen. Andere Vorgehensweisen, wie z.B. die SchŸlerexperimente vorfŸhren zu lassen, erscheinen der Lehrperson nicht objektiv genug. Letztendlich wissen die SchŸler jedoch selbst am besten, was zum VerstŠndnis am meisten beigetragen hat, weshalb deren Meinung und EindrŸcke berŸcksichtigt werden mŸssen. Deshalb ist besonderer Wert auf Feedback gelegt worden, auch um das weitere Vorgehen daraufhin optimieren zu kšnnen.

\subsection{Beschreibung der Lerngruppe}
Die Lerngruppe besteht aus vier SchŸlerinnen und neun SchŸlern. Es handelt sich um einen S1/S2 Grundkurs Physik, den ich im Schuljahr 2008/2009 selbststŠndig unterrichtet habe. An der Kooperativen Schule Tonndorf sind fŸr Physik in dieser Jahrgangsstufe drei Stunden pro Woche vorgesehen. Hiervon sind zwei Stunden zu einer Doppelstunde zusammengefasst worden. Bedingt durch die Umstellung von G9 auf G8 sind sowohl SchŸler aus der 10. wie aus der 11. Klasse in den S1 Grundkurs Physik gekommen. Tendenziell sind letztere deutlich leistungsstŠrker. Zwei SchŸler haben das letzte Schuljahr in SŸdamerika verbracht und die Klasse Ÿbersprungen mit der Konsequenz, dass der Lernstand des einen SchŸlers sich Ÿber dem Kursniveau bewegt, wŠhrend der andere SchŸler deutliche Defizite aufweist. Drei SchŸler (Alina, Steven, Daniel) kšnnten problemlos in einen Leistungskurs wechseln, welcher an der Kooperativen Schule Tonndorf jedoch nicht hat eingerichtet werden kšnnen. 
Die in der Mittelstufe erworbene Sach- und Methodenkompetenz vieler SchŸler entspricht, aus mir nicht bekannten GrŸnden, nicht den zu erwartenden Vorkenntnissen. Drei SchŸler (Per, Ahmad, Markus) haben Physik bereits fŸr das nŠchste Halbjahr abgewŠhlt, was sich auch in deren Motivation widerspiegelt, da sie sich manchmal einer aktiven Mitarbeit entziehen wollen. In persšnlichen GesprŠchen haben sie erwŠhnt, dass sie sich das Fach anders vorgestellt und in der Mittelstufe keinen wirklichen Eindruck vom Fach gewonnen haben. Diesen Widrigkeiten zum Trotz zeigt der Rest der Lerngruppe gro§es Interesse, allerdings manchmal mehr an LehrervortrŠgen, als an selbst erarbeiteten Erkenntnissen. So zeigen Artur, Pascal und Quoc-Thai gro§es Interesse an den bedeutenden physikalischen Fragestellungen wie RelativitŠt, lassen sich diese gerne ausfŸhrlich erklŠren und fragen bei VerstŠndnisschwierigkeiten nach. Auch bei der eigenstŠndigen Auseinandersetzung mit physikalischen Problemen arbeiten sie zufriedenstellend mit, jedoch deutlich passiver. Alina und Friederike dagegen sind besonders dann hoch motiviert, wenn sie selber eigenstŠndig z. B. an einem Versuch arbeiten. Ihr Enthusiasmus ist deutlich erkennbar, wenn ihre Arbeit erfolgreich verlaufen ist. Alina, Daniel und Steven sind die leistungsstŠrksten SchŸler im Kurs. Sowohl ihre Mitarbeit im UnterrichtsgesprŠch, als auch ihre TŠtigkeiten bei selbststŠndig zu bearbeitenden Aufgaben bzw. Experimenten ist gut. Allerdings mŸssen auch Daniel und Steven manchmal daran erinnert werden, gewisse Aufgaben, die sie als lŠstig ansehen, gewissenhaft durchzufŸhren. Janna, Asgard und Franziska sind gute SchŸler, die im UnterrichtsgesprŠch etwas zurŸckhaltend sind; wŠhrend einer Gruppenarbeitsphase stellt sich jedoch schnell heraus, dass sie gut, produktiv und zielfŸhrend zusammenarbeiten. 
Zu Beginn des zweiten Halbjahres sind die SchŸler befragt worden, ob sie einer Umstellung der Arbeitssprache auf Englisch begrŸ§en wŸrden, da die unterrichtende Lehrkraft lange Zeit in den Vereinigten Staaten gearbeitet hat und sich dieses daher zutrauen wŸrde. Die Resonanz auf dieses Angebot ist durchweg positiv gewesen, weshalb das zweite Halbjahr durchgŠngig bilingual unterrichtet worden ist. Die Kommunikation mit den SchŸlern bereitet weder auf Englisch noch auf Deutsch Probleme.

\subsection{Sachanalytische Hinweise}
WiderstŠnde und Kondensatoren gehšren zu den grundlegenden passiven elektronischen Bauteilen. Sie finden sich in gro§er Zahl auf Platinen elektronischer GerŠte oder integriert in Mikrochips. Jeder "normale"' elektrische Leiter besitzt einen Widerstand, der in AbhŠngigkeit vom Spannungsabfall Ÿber die LŠnge des Leiters die Hšhe des Stromflusses bestimmt. Hierbei wird ein Teil der Energie in WŠrme umgesetzt. Zur †bertragung elektrischer Energie ist dies nachteilig; in elektronischen Schaltungen kann dies jedoch ausgenutzt werden, um den Stromfluss zu begrenzen. Hierzu stellt die Industrie WiderstŠnde (das Bauteil) in einer Vielzahl verschiedener Widerstandswerte her.
Ein Kondensator besteht im einfachsten Fall aus zwei leitenden parallelen Platten. Legt man zwischen ihnen eine Spannung an, so pumpt die Spannungsquelle Elektronen von der einen auf die andere Platte, wodurch sich auf der einen Platte ein LadungsŸberschuss und auf der anderen Platte ein betragsmŠ§ig gleich gro§er Ladungsmangel einstellt. Die Grš§e der Ladung im stationŠren Zustand hŠngt von der angelegten Spannung und von der KapazitŠt des Kondensators ab, welche im Falle des Plattenkondensators u.a. von der Grš§e und dem Abstand der Platten bedingt ist. Vergrš§ert man den Abstand, so sinkt die KapazitŠt (alternativ kšnnen zwei Kondensatoren in Serie geschaltet werden). Vergrš§ert man die FlŠche der Platten, so vergrš§ert sich die KapazitŠt (alternativ kšnnen zwei Kondensatoren parallel geschaltet werden). Das Pumpen der Ladungen erfordert Energie, da sowohl beim Entfernen einer negativen Ladung von der nun positiv geladenen Platte, als auch beim Auftragen dieser Ladung auf die nunmehr negativ geladenen Platte, eine Kraft Ÿberwunden werden muss. Diese Energie geht nicht verloren, sondern ist im elektrischen Feld, welches sich aufgrund der Ladung der Platten entwickelt, gespeichert. Beim Aufladen des Kondensators ergibt sich ein charakteristischer Zusammenhang zwischen der Ladezeit und der gespeicherten Ladung. Ist der Kondensator aufgeladen, kommt der Stromfluss zum Erliegen (was theoretisch erst nach unendlich langer Zeit der Fall ist). Beim Anlegen einer Wechselspannung ist jedoch ein stationŠrer Stromfluss mšglich, da der Kondensator immer fŸr kurze Zeit auf- bzw. entladen wird. Bei unendlich hohen Frequenzen stellt der Kondensator daher einen Kurzschluss dar und bei Gleichspannung im stationŠren Fall einen unendlich gro§en Widerstand. 
Kondensatoren kšnnen daher als Filter eingesetzt werden, um bestimmte Frequenzanteile zu blocken oder durchzulassen. Mit dem Aufkommen von Kondensatoren sehr gro§er KapazitŠt in der Grš§enordnung von Farad wird auch die Verwendung zur Speicherung elektrischer Energie immer interessanter.
WiderstŠnde bzw. Kondensatoren kšnnen parallel und in Serie geschaltet werden. Serienschaltung zweier WiderstŠnde fŸhrt zu einem hšheren Widerstand, Parallelschaltung zu einem verminderten Widerstand. FŸr Kondensatoren gilt das Umgekehrte.

\subsection{(P)SPICE}
SPICE (Simulation Program with Integrated Circuit Emphasis) ist ein Programm zur Simulation analoger elektronischer Schaltungen. Die erste Version von SPICE ist im Electronics Research Laboratory an der University of California, Berkeley entwickelt und 1973 der …ffentlichkeit vorgestellt worden. Der Bedarf nach einem solchen Programm hat sich durch die zunehmende KomplexitŠt integrierter Schaltungen ("`Mikrochips"') ergeben. Ein Testaufbau der Schaltung mit diskreten Bauteilen auf einer Platine sagt letztendlich Ÿber die FunktionsfŠhigkeit der gleichen Schaltung, die nun auf einem Mikrochip untergebracht werden soll, wenig aus. Ein Grund hierfŸr sind z. B. parasitŠre KapazitŠten (KapazitŠten, die unbeabsichtigt entstehen, weil z. B. zwei Leiterbahnen sehr dicht beieinander liegen), die sich zwangslŠufig bei den winzigen Abmessungen auf einem Chip ergeben. Solche Stšreffekte lassen sich einfacher mit einem Computerprogramm als mit einem echten Schaltungsaufbau modellieren. Trotzdem kann SPICE (bzw. die PC-Version PSPICE \cite{pspice91} \cite{orcad}) natŸrlich auch zur Simulation einfacherer Schaltungen verwendet werden, die mit diskreten Bauteilen z. B. auf einer Platine aufgebaut werden. Die Schaltungssimulation mit (P)SPICE kann als Standard in der Industrie angesehen werden. 
€ltere Versionen machen zur Eingabe des Schaltkreises von einer Netzliste Gebrauch. Hierzu wird ein Schaltplan auf ein StŸck Papier gezeichnet und jeder Knotenpunkt mit einer Nummer versehen. Die Lage eines Bauteils kann nun angegeben werden indem die Nummer des Knotenpunktes fŸr jeden Anschluss, der mit ihm verbunden ist, eingegeben wird. Eine typische Zeile in einer SPICE-Eingabedatei kšnnte wie folgt aussehen:

R1 3 0 1k

Dies bedeutet, dass sich ein Widerstand mit der Bezeichnung "`1"' und dem Widerstandswert $1k\Omega$ zwischen den Knoten 3 und 0 befindet. †blicherweise ist der Knotenpunkt 0 reserviert fŸr die Masse. Alle Potentiale werden in Bezug auf diesen Knoten errechnet. Weiterhin stehen Befehle zur VerfŸgung, die den Betriebsmodus des Programms festlegen. So kann SPICE z.B. im Transientenmodus arbeiten (d. h. die ZeitabhŠngigkeit aller Grš§en wird errechnet), im AC-Modus (alle Grš§en werden im stationŠren Wechselspannungsbetrieb berechnet) oder im DC-Modus (dito fŸr Gleichspannung). 
Neuere Versionen von SPICE ermšglichen auch die grafische Eingabe des Schaltplans sowie die grafische Ausgabe der berechneten Grš§en.

\subsection{Didaktische Entscheidungen}
So gut wie die gesamte Lerngruppe langweilt sich bei der Bearbeitung von Standardproblemen und bei der Erledigung von Routineaufgaben, weshalb auch Hausaufgaben hŠufig nicht gemacht werden. So sollte in einer Hausaufgabe, die besonders viel Routinearbeit nach sich fŸhrt, die Periodendauer eines aus KŸchenutensilien selbstgebauten harmonischen Oszillators mit Hilfe einer Stoppuhr bestimmt werden. Diese Aufgabe ist nur von einem einzigen SchŸler gemacht worden. Die SchŸler finden Gefallen an den modernen Problemen der Physik wie Quantenmechanik oder RelativitŠtstheorie, doch diese Gebiete kšnnen noch nicht behandelt werden. Die mŸndliche Mitarbeit beschrŠnkt sich bei ein paar SchŸlern wie Per, Ahmad und Markus auf (durchaus) interessiertes Zuhšren; eine aktivere Mitarbeit ist mit klassischen †bungsaufgaben nur schwer zu erzielen.
Die grobe Zielrichtung in der hier vorliegenden Unterrichtseinheit besteht darin, die physikalischen Grundlagen a.) interessanter Probleme aus der Technik von den SchŸlern b.) unter Abnahme von Routinearbeit c.) selbststŠndig entdecken zu lassen.
Das der Unterrichtseinheit zugrunde liegende Konzept kšnnte als exploratives Experimentieren bezeichnet werden und wird in \cite{Aufschnaiter:2008sh} beschrieben. Das Gegenteil hiervon ist das "`hypothesenprŸfende Experimentieren"', welches hŠufig im Physikunterricht anzutreffen ist. Bei letzterem werden exemplarisch nur einige Versuche durchgefŸhrt, deren Sinn in der Verifizierung einer zuvor behandelten Theorie oder zur Motivation einer zu erarbeitenden Theorie liegt. Sie haben prototypischen Charakter und schlie§en Irrwege im Vorweg aus. Die Sinn- und Zweckhaftigkeit eines solchen Vorgehens wird in der Literatur mittlerweile angezweifelt \cite{Aufschnaiter:2008sh}, weil die SchŸler im ersten Fall nicht Ÿber genug Theoriekenntnisse oder im zweiten Fall nicht Ÿber genug Beispiele verfŸgen, um die relevanten Parameter zu erkennen und Hypothesen ableiten zu kšnnen. Beim explorativen Experimentieren hingegen wird der Einfluss von Parameter durch deren systematische Variation untersucht. Dies ist zeitaufwendig und erfordert langwierige Experimentierserien, kommt jedoch der wirklichen Forschung wesentlich nŠher. SchŸler und Wissenschaftler befinden sich hŠufig in der gleichen Situation, wenn ihnen die zugrundeliegende Theorie nicht bekannt ist. In einem solchen Fall werden auch Experimentalphysiker vorsichtig den Einfluss jedes denkbaren Parameters untersuchen. Viele Parameter werden sich dabei fŸr die Problemstellung als irrelevant erweisen. Jedoch dŸrfen auch SchŸler dieser Erfahrung nicht beraubt werden, wenn sie in der Lage sein sollen, aufgrund von Beobachtungen induktiv auf allgemeine ZusammenhŠnge zu schlie§en. 
Die Lehrkraft sieht sich hierbei nur als Begleiter, die den SchŸlern beratend zur Seite steht und Anregungen fŸr eigene Experimente liefert und darauf achtet, dass sich keine Fehlvorstellungen einschleichen, die spŠter nur schwierig zu korrigieren sind.
Laut Lehrplan \cite{rahmenplan} sollen im S1/S2-Grundkurs Physik Eigenschaften und Anwendungen des Kondensators behandelt werden. Hierzu sind Grundkenntnisse aus der Mittelstufe Ÿber das Ohmsche Gesetz und die Kirchhoffschen Regeln erforderlich, die bei vielen SchŸlern jedoch nicht vorhanden gewesen sind. Ferner fordert der Rahmenplan einen zunehmenden Einsatz von Arbeitsformen, die verstŠrkt an ein eigenverantwortliches Lernen heranfŸhren. Die SchŸler sollen hierbei die Gelegenheit bekommen, ihren Lernprozess im Wesentlichen selbst zu steuern \cite{altmann:2007ve} \cite{Gudjons:2001ly}.
Der Rahmenplan sieht den Erwerb von Kompetenzen in den Bereichen "`Kenntnisse"', "`Fachmethoden"', "`Kommunikation"' und "`Kontexte"' vor. In der hier vorliegenden Unterrichtseinheit sollen die folgenden Bereiche besonders gefšrdert werden:

\subsubsection{Kenntnisse}
Wissen aneignen (durch Schlussfolgerungen aus beobachteten Versuchen)

\subsubsection{Fachmethoden}
Methoden des Experimentierens (Planung, DurchfŸhrung und Auswertung von Experimenten)
Strategien der Erkenntnisgewinnung (durch Experimentierserien)

\subsubsection{Kommunikation}
Methoden der Darstellung (durch Benutzung von Tabellen und Diagrammen zur Darstellung der Ergebnisse)

\subsubsection{Kontexte}
Beziehung zwischen Physik und Technik (Einsatz von Computersimulationen in der Industrie, Einsatz von Grundschaltungen in technischen Anwendungen)

\subsection{Ziele der Unterrichtseinheit}
Ziel der Unterrichtseinheit ist es, den SchŸlern die Grundlagen elektrischer Schaltkreise zu vermitteln. Dazu gehšrt die Unterscheidung zwischen Spannung und Strom, das Ohmsche Gesetz, die Kirchhoffschen Regeln und der Zusammenhang zwischen Ladung, KapazitŠt und Spannung bei Kondensatoren. Die SchŸler kšnnen selbststŠndig Experimente entwerfen, durchfŸhren und insbesondere Spannungen und Stršme messen. Die gewonnenen Ergebnisse kšnnen sie fachsprachlich korrekt aufzeichnen und wiedergeben.

\section{DurchfŸhrung}
\subsection{Unterrichtsverlauf}
Der Arbeitsauftrag hat daher darin bestanden, Schaltungen mit WiderstŠnden und spŠter auch mit Kondensatoren aufzubauen und zu variieren. Um den Lernfortschritt der SchŸler zu ermitteln, ist vor jeder Experimentiereinheit ein Test geschrieben worden, der den Ist-Zustand der vorhandenen Kenntnisse hat ermitteln sollen. Diesen Tests ist eine doppelte Funktion zugekommen, da sie au§erdem den SchŸlern Variationsmšglichkeiten aufzeigen sollten, die als Ausgangspunkt fŸr eigene Versuche haben dienen kšnnen. 
Aufgrund der Untersuchungsfrage ist die Mšglichkeit der grafischen Eingabe des Schaltplans nicht zum Einsatz gekommen. Jedoch gibt es hierfŸr auch didaktische GrŸnde. So zwingt die Eingabe des Schaltplans Ÿber eine Netzliste zur genauen Protokollierung der Versuche, da fŸr jede Schaltung ein Schaltplan gezeichnet werden muss und die Knotenpunkte nummeriert werden mŸssen. Dies resultiert in eine bewusstere Planung und Auseinandersetzung mit den Experimenten. Ein weiterer Grund liegt in der MŠchtigkeit des Simulationsprogramms, dessen Funktionsumfang den UngeŸbten leicht Ÿberfrachtet. Die Bauteilbibliothek von PSPICE umfasst tausende von Modellen, von denen nur sehr wenige benštigt werden. Das gleiche gilt fŸr die zur VerfŸgung stehenden Analysearten, von denen nur ein paar im Unterricht behandelt werden kšnnen. StŸnde der volle Funktionsumfang durch die blo§e Auswahl der Teile aus einer langen Liste zur VerfŸgung, so kŠme dies dem Auspacken sŠmtlicher im Fachbereich vorhandenen GerŠte gleich. Durch die Benutzung von Netzlisten kšnnen die benštigten Befehle nach Bedarf bekannt gegeben werden. Somit erfolgt eine Fokussierung auf die fŸr die Unterrichtseinheit relevanten Probleme. 
Die DurchfŸhrung des Unterrichtsversuchs hat sich Ÿber einen Zeitraum von etwa sechs Wochen erstreckt. Physik wird im S2-Grundkurs dreistŸndig unterricht, wobei zwei Stunden zu einer Doppelstunde zusammengefasst sind. Innerhalb des Untersuchungszeitraums sind die Doppelstunden immer fŸr Experimente bzw. Computersimulationen genutzt worden. In den Einzelstunden sind neue Konzepte eingefŸhrt und Fragen bzw. Probleme behandelt worden, die fŸr den gesamten Kurs von Interesse gewesen sind. Mit 13 SchŸlern ist die Kursgrš§e leider sehr niedrig, sodass statistisch relevante Aussagen kaum zu treffen sind. Erschwert worden ist der Unterrichtsversuch durch das hŠufige Fehlen von SchŸlern aus privaten / gesundheitlichen GrŸnden oder aufgrund von anderen schulischen Verpflichtungen. Dadurch bedingt gibt es nur wenige vollstŠndige DatensŠtze. 
Der gesamte Unterricht einschlie§lich aller Tests, Klausuren und Arbeitsmaterialien ist in englischer Sprache erfolgt. Die besonderen Erfordernisse des bilingualen Physikunterrichts sollen hier jedoch nicht wiedergegeben werden. Es gibt keinerlei Anzeichen dafŸr, dass die englische Arbeitssprache die Ergebnisse in irgendeiner Weise verfŠlscht hŠtten. FŸr den "`Notfall"' haben immer entsprechende Hilfsmittel bereit gestanden. In regelmŠ§igen Feedback-Runden haben die SchŸler immer wieder betont, dass die englische Sprache keine zusŠtzliche HŸrde darstellt. Da viele dieser Runden kein anonymes Feedback ermšglicht haben, ist es nicht offensichtlich, dass sich die SchŸler tatsŠchlich frei geŠu§ert haben. AnfŠnglich sind die SchŸler in ihren €u§erungen auch sehr zšgerlich gewesen; spŠter jedoch sind die GesprŠche als sehr offen zu bezeichnen gewesen, so dass es keinen Grund gibt, an der Korrektheit der Aussagen zu zweifeln. Die Ergebnisse unterscheiden sich auch nicht von denen vom Feedback-Bogen, welcher anonym erhoben worden ist, ohne nach den Namen der SchŸler zu fragen. Viele SchŸler haben allerdings trotzdem freiwillig ihren Namen preisgegeben.

Thematisch behandelt worden sind die Gebiete "Anordnung von Strom- bzw. Spannungsquellen"', Anordnung von WiderstŠnden und Gleichstromverhalten des Kondensators. Bei den ersten beiden Themengebieten hat es sich um reine Wiederholung von Mittelstufenkenntnissen gehandelt. Die Wiederholung hat die Gelegenheit geboten, dass sich SchŸler mit dem Simulationsprogramm bzw. den ExperimentierkŠsten mit Hilfe eines nicht všllig unbekannten Themas vertraut machen. Auch ist es so den SchŸlern ermšglicht worden, inhaltlich auf einen gemeinsamen Nenner zu kommen.  

\subsubsection{Quiz \#1} 
Quiz \#1 besteht aus fŸnf Aufgaben zum Thema Spannungsquellen. Gefragt ist immer nach dem Gesamtstrom, wobei die Last immer aus genau einem Widerstand besteht. In der ersten Ausgabe gibt es nur eine Batterie und in der zweiten Aufgabe zwei in Serie geschaltete Batterien als Quelle. Aufgabe 3 und 4 unterscheiden sich von der zweiten Aufgabe nur darin, dass die Batterien gegensŠtzlich gepolt sind und die Spannungen daher subtrahiert werden mŸssen. Zum Schluss sind in der letzten Aufgabe zwei Batterien mit gleicher Spannung parallel geschaltet. Auf die InnenwiderstŠnde der Batterien wird an keiner Stelle eingegangen.

\subsubsection{Quiz \#2}
Quiz \#2 ist eine exakte Kopie von Quiz \#1. Dadurch sollte den SchŸlern gezeigt werden, dass im Prinzip alle Aufgaben durch die sorgfŠltige DurchfŸhrung der Versuche / Simulationen lšsbar sind.

\subsubsection{Quiz \#3}
In Quiz \#3 geht es um die Parallel- und Serienschaltung von WiderstŠnden.  Als Spannungsquelle kommt immer genau eine Batterie zum Einsatz. Gefragt wird sowohl nach dem Gesamtstrom als auch nach SpannungsabfŠllen Ÿber einzelne WiderstŠnde. 
\begin{itemize}
\item Aufgabe 1: ein Widerstand

\item Aufgabe 2:  zwei WiderstŠnde in Serie

\item Aufgabe 3: Parallelschaltung zweier WiderstŠnde

\item Aufgabe 4: Serienschaltung eines einzelnen und zweier parallel geschalteter WiderstŠnde
\end{itemize}
\subsubsection{Quiz \#4}
Thematisch geht es Ð wie in Quiz \#3 - um die Parallel- und Serienschaltung von WiderstŠnden. Allerdings sind die Werte und die Reihenfolge der Aufgaben aus Quiz \#3 verŠndert worden. Teilweise sind die Schaltbilder bei gleicher Schaltung anders gezeichnet worden. 
\begin{itemize}
\item Aufgabe 1: entspricht Aufgabe 1 aus Quiz \#3

\item Aufgabe 2: entspricht Aufgabe 3 aus Quiz \#3

\item Aufgabe 3: entspricht Aufgabe 3 aus Quiz \#3

\item Aufgabe 4: entspricht Aufgabe 2 aus Quiz \#3

\item Aufgabe 5: entspricht Aufgabe 4 aus Quiz \#3, jedoch werden nun vier WiderstŠnde, von denen jeweils zwei in parallel geschaltet sind, in Serie geschaltet
\end{itemize}
\subsubsection{Quiz \#5}
(nicht Teil des Unterrichtsversuchs)

\subsubsection{Quiz \#6}
In Quiz \#6 wird Wissen Ÿber  Kondensatoren abgefragt.
\begin{itemize}
\item Aufgabe 1: Wird die gespeicherte Ladung beim Anlegen einer hšheren Spannung grš§er? Da die Wahrscheinlichkeit dafŸr, die richtige Antwort zu erraten, bei 50\% liegt, wird die Aufgabe in der Auswertung nicht berŸcksichtigt. Die Aufgabe dient lediglich zum "`Warmwerden"'.

\item Aufgabe 2: Hier soll der Zusammenhang Q=CU angewendet werden

\item Aufgabe 3: Verhalten der KapazitŠt eines Plattenkondensators beim Auseinanderziehen der Platten. Es gilt die gleiche EinschrŠnkung wie in Aufgabe 1.

\item Aufgabe 4: Verhalten der KapazitŠt eines Plattenkondensators beim Vergrš§ern der Platten. Es gilt die gleiche EinschrŠnkung wie in Aufgabe 1.

\item Aufgabe 5: Qualitatives und quantitatives Verhalten bei Reihenschaltung zweier Kondensatoren

\item Aufgabe 6: Qualitatives und quantitatives Verhalten bei Parallelschaltung zweier Kondensatoren
\end{itemize}
\subsubsection{Quiz \#7}
Quiz \#7 ist eine Kopie von Quiz \#6 mit verŠnderten Werten.

\subsubsection{Final Exam}
Der Wiederholungsteil der letzten Klausur ist eine Kopie von Quiz \#6 bzw. Quiz \#7 mit anderen Werten. Da sich die meisten SchŸler auf die Klausur speziell vorbereitet haben, ist die Aussagekraft der Ergebnisse begrenzt. 

Um zu sehen, ob die SchŸlerexperimente bzw. die Computersimulationen sich auf das physikalische VerstŠndnis positiv auswirken, ist jeweils vor und nach einer Experimentier- bzw. Simulationseinheit ein Quiz zum Thema geschrieben worden. Den SchŸlern ist gesagt worden, dass es sich beim hier vorliegenden Projekt um eine reine Forschungsangelegenheit handele und die Ergebnisse der Quiz nicht in die Note einflie§en wŸrden. Dies ist einerseits notwendig gewesen, um die SchŸler in Anbetracht der hohen Anzahl an Tests nicht zu verschrecken. Dies wŠre auch deswegen unfair gewesen, weil die genauen Vorkenntnisse der SchŸler aus der Mittelstufe stark variieren. 
Eine Bewertung des korrespondieren Tests nach dem Ende der Experimentiereinheit wŠre denkbar gewesen. Jedoch hŠtten sich die SchŸler dann womšglich viel stŠrker zu Hause auf die entsprechenden Fragen vorbereit, wodurch eine Korrelation der Testergebnisse mit der eigentlichen Fragestellung unzulŠssig wŠre. Problematisch bei diesem Vorgehen ist natŸrlich die Gefahr, dass sich SchŸler bei der Bearbeitung der Quiz eventuell nicht besonders sorgfŠltig gearbeitet haben (Ein sofortiges Abgeben eines offensichtlich nicht bearbeiteten Testbogens wŠre jedoch als eine ungenŸgende mŸndliche Mitarbeit angesehen worden). 

Folgende chronologische Abfolge hat sich ergeben:

{\raggedright

\vspace{3pt} \noindent
\begin{tabular}{|p{144pt}|p{144pt}|p{144pt}|}
\hline
\parbox{144pt}{\raggedright 
{\small Datum}
} & \parbox{144pt}{\raggedright 
{\small Arbeitsauftr\"{a}ge}
} & \parbox{144pt}{\raggedright 
{\small Thema}
} \\
\hline
\parbox{144pt}{\raggedright 
{\small 7.5.09}
} & \parbox{144pt}{\raggedright 
{\small Quiz \#1}

{\small Experiment / Simulation}

} & \parbox{144pt}{\raggedright

{\small Spannungsquellen}
} \\
\hline
\parbox{144pt}{\raggedright 
{\small 14.5.09}
} & \parbox{144pt}{\raggedright 
{\small Quiz \#2}

{\small Quiz \#3}

{\small Simulation / Experiment}
} & \parbox{144pt}{\raggedright

{\small Widerst\"{a}nde}
} \\
\hline
\parbox{144pt}{\raggedright 
{\small 28.5.09}
} & \parbox{144pt}{\raggedright 
{\small Quiz \#4}
} & \parbox{144pt}{\raggedright 
{\small Widerst\"{a}nde}
} \\
\hline
\parbox{144pt}{\raggedright 
{\small 4.6.09}
} & \parbox{144pt}{\raggedright 
{\small Quiz \#6}

{\small Experiment / Simulation}
} & \parbox{144pt}{\raggedright

{\small Kondensatoren}
} \\
\hline
\parbox{144pt}{\raggedright 
{\small 11.6.09}
} & \parbox{144pt}{\raggedright 
{\small Quiz \#7}
} & \parbox{144pt}{\raggedright 
{\small Kondensatoren}
} \\
\hline
\end{tabular}
\vspace{2pt}

}

Tests sind also nie im direkten Anschluss in der gleichen Stunde nach einer Experimentier- bzw. Simulationseinheit geschrieben worden, da es vermutlich schwierig gewesen wŠre, die SchŸler zu einem solchen Wechsel der Arbeitsform zu bewegen. 
Die SchŸler sind in zwei Gruppen eingeteilt worden, die entweder reale Experimente durchgefŸhrt haben oder die entsprechenden Experimente am Computer simuliert haben. Bei der zweiten Einheit sind die Gruppen einfach vertauscht worden. Eine vierte Einheit hat bedauerlicherweise nicht stattfinden kšnnen, da aufgrund von gesundheitlichen GrŸnden und anderen schulischen Verpflichtungen bei einer Doppelstunde weniger als die HŠlfte der 13 sich im Kurs befindlichen SchŸler anwesend gewesen sind. Daher sind in der letzten Einheit einfach zwei ungefŠhr gleich gro§e neue Gruppen gebildet worden. 
Alle Versuche am Computer und mit realen Bauteilen sind von den SchŸlern dokumentiert worden. Insbesondere ist viel Wert darauf gelegt worden, dass die SchŸler genaue Angaben Ÿber jeden neuen Versuch, jede VerŠnderung eines Versuchs sowie die dafŸr benštigte Zeit notieren. 
Das Experimentierverhalten ist in jeder Doppelstunde mit einer Videokamera festgehalten worden. 
Am Ende des Unterrichtsversuchs haben die SchŸler ihre Erfahrungen auf einem RŸckmeldebogen zu Papier gebracht. Nebst allgemeinen Fragen nach VerbesserungsvorschlŠgen etc. ist speziell gefragt worden, welche Vor- und Nachteile sie jeweils bei der Verwendung realer Experimente bzw. Computersimulationen sehen und welche Probleme bei beiden Methoden aufgetaucht sind. Au§erdem ist gefragt worden, ob sie eine begrŸndete PrŠferenz fŸr eine Vorgehensweise haben und welche ihrer Meinung nach den grš§ten Lernerfolg verspricht. 

\subsection{Die Simulationen}
Dem Fachbereich Physik stehen sieben Laptops zur VerfŸgung, auf denen das Schaltungssimulationsprogramm PSPICE in der Version 9.1 installiert worden ist. Das Programm kann entweder grafisch durch das Zeichnen eines Schaltplans bedient werden oder "`klassisch"' durch das Bearbeiten einer Textdatei, in der entsprechende Befehle zur Erstellung einer Netzliste eingefŸgt werden. Aufgrund der zu untersuchenden Fragestellung ist die Wahl auf die letztere Betriebsart gefallen. Eine †bersicht mit den relevanten Befehlen ist verteilt worden. Ebenfalls sind Beispieldateien und eine einseitige Kurzanleitung ausgeteilt worden, mit denen ausgehend von einem Schaltbild die entsprechende Eingabedatei erstellt werden kann.
Je nach Einstellung arbeitet das Programm im DC-, AC- oder im Transientenmodus. Im DC-Modus wird das stationŠre Gleichspannungsverhalten ermittelt. Dazu schreibt das Programm alle berechneten Stršme sowie alle Spannungen an jedem Knotenpunkt (in Bezug auf einen speziellen Knoten) in eine Ausgabedatei. Im Transientenmodus wird zusŠtzlich die ZeitabhŠngigkeit der einzelnen Grš§en berŸcksichtigt. Dazu kšnnen alle eben genannten Spannungen und Stršme vom Programm grafisch  als eine Funktion der Zeit dargestellt werden. FŸr die Untersuchung des Lade- bzw. Entladevorgangs von Kondensatoren ist der Transientenmodus herangezogen worden. Ansonsten genŸgt fŸr die hier behandelten Probleme der DC-Modus.

\subsection{Die Experimente}
Ebenfalls zur VerfŸgung stehen acht ElektronikexperimentierkŠsten sowie entsprechende Netzteile. Au§erdem existieren 4,5V-Kastenbatterien und Multimeter. Die letzten beiden GegenstŠnde sind jedoch hŠufig nicht in genŸgend gro§er Anzahl vorhanden gewesen, da zahlreiche Multimeter defekt sind (typischerweise sind hier defekte Feinsicherungen das Problem) und auch viele Batterien nicht verwendet haben werden kšnnen (leere oder ausgelaufene Batterien bzw. Batterien mit defekten AnschlŸssen). Da einige Experimente gleich mehrere Batterien bzw. MessgerŠte erfordern, ist es trotz der geringen SchŸlerzahl zu Wartezeiten gekommen. 
Zusammenfassend kann der Verlauf einer Doppelstunde wie folgt beschrieben werden: Die Stunde beginnt mit zwei Quiz, welche ungefŠhr 20 Minuten in Anspruch nehmen. Das erste Quiz fragt thematisch die Themen aus der letzten Doppelstunde ab (au§er in der ersten Doppelstunde der Unterrichtseinheit). Das zweite Quiz dient der Erfassung des Leistungsstandes fŸr die in Doppelstunde behandelten Themen. Die Quiz sind von den SchŸlern sehr diszipliniert bearbeitet worden. Danach erfolgt eine Aufteilung der Lerngruppe in zwei Gruppen, von denen die eine reale Experimente und die andere Computersimulationen zum Thema durchfŸhrt.

\section{Auswertung}
\subsection{Tests}

{\raggedright

\vspace{3pt} \noindent
\begin{tabular}{|p{40pt}|p{76pt}|p{76pt}|p{98pt}|p{80pt}|}
\hline
\parbox{40pt}{\raggedright 
\textbf{{\small Aufgabe}}
} & \parbox{76pt}{\raggedright 
\textbf{{\small Verbessert (exp)}}
} & \parbox{76pt}{\raggedright 
\textbf{{\small Verbessert  (sim)}}
} & \parbox{98pt}{\raggedright 
\textbf{{\small Verschlechtert (exp)}}
} & \parbox{80pt}{\raggedright 
\textbf{{\small Verschlechtert (sim)}}
} \\
\hline
\parbox{40pt}{\raggedright 
{\small 1}
} & \parbox{76pt}{\raggedright 

} & \parbox{76pt}{\raggedright 
{\small 2}
} & \parbox{98pt}{\raggedright 

} & \parbox{80pt}{\raggedright 

} \\
\hline
\parbox{40pt}{\raggedright 
{\small 2}
} & \parbox{76pt}{\raggedright 

} & \parbox{76pt}{\raggedright 
{\small 2}
} & \parbox{98pt}{\raggedright 

} & \parbox{80pt}{\raggedright 

} \\
\hline
\parbox{40pt}{\raggedright 
{\small 3}
} & \parbox{76pt}{\raggedright 

} & \parbox{76pt}{\raggedright 
{\small 3}
} & \parbox{98pt}{\raggedright 
{\small 2}
} & \parbox{80pt}{\raggedright 
{\small 1}
} \\
\hline
\parbox{40pt}{\raggedright 
{\small 4}
} & \parbox{76pt}{\raggedright 
{\small 1}
} & \parbox{76pt}{\raggedright 
{\small 2}
} & \parbox{98pt}{\raggedright 
{\small 1}
} & \parbox{80pt}{\raggedright 
{\small 1}
} \\
\hline
\parbox{40pt}{\raggedright 
{\small 5}
} & \parbox{76pt}{\raggedright 

} & \parbox{76pt}{\raggedright 
{\small 3}
} & \parbox{98pt}{\raggedright 
{\small 1}
} & \parbox{80pt}{\raggedright 

} \\
\hline
\parbox{40pt}{\raggedright 
{\small gesamt}
} & \parbox{76pt}{\raggedright 
{\small 1}
} & \parbox{76pt}{\raggedright 
{\small 12}
} & \parbox{98pt}{\raggedright 
{\small 4}
} & \parbox{80pt}{\raggedright 
{\small 2}
} \\
\hline
\end{tabular}
\vspace{2pt}

}
Interessanterweise kommt es also auch zu Verschlechterungen. BerŸcksichtigt man die Mšglichkeit einer Verschlechterung, so kšnnen sich die Computersimulationen gegenŸber den gewšhnlichen SchŸlerexperimenten behaupten. Allerdings muss darauf hingewiesen werden, dass sich Verbesserungen bzw. Verschlechterungen bei einigen SchŸlern hŠufen (Beispiel: Artur hat im ersten Quiz alle Fragen falsch beantwortet und im zweiten Quiz alle Frage richtig. In den obigen Tabellen wŸrde dies als fŸnf Verbesserungen ausgewiesen werden.). Daher wird der beobachtete Effekt durch die ZŠhlweise getrennt nach Aufgaben verstŠrkt und ist daher mšglicherweise wesentlich schwŠcher. 
Da es bei den Tests relativ wenig Punkte gibt und im Wesentlichen nur nach richtig oder falsch bewertet wird, was bei den hier vorliegenden Tests sehr einfach mit hoher Genauigkeit mšglich ist, ist die Aufteilung in SchŸler, die sich verbessert bzw. verschlechtert haben, sinnvoll (gŠbe es sehr viele Punkte, so wŠre es sehr unwahrscheinlich, dass ein SchŸler genau die gleiche Punktzahl erreicht. Selbst bei gleicher Leistung des SchŸlers mŸsste dann die Ungenauigkeit bei der Bewertung berŸcksichtigt werden.).

\subsection{Protokolle}

Anhand der Versuchsprotokolle ist ermittelt worden, wie viele Versuche (einschlie§lich Variationen) insgesamt durchgefŸhrt worden sind. Ebenfalls ausgewertet worden sind die Zeitangaben pro Experiment / Simulation der SchŸler.

{\raggedright

\vspace{3pt} \noindent
\begin{tabular}{|p{92pt}|p{83pt}|p{74pt}|p{83pt}|p{74pt}|}
\hline
\parbox{92pt}{\raggedright 

} & \parbox{83pt}{\raggedright 
\textbf{{\small Experimente (Spannungs-quellen und Widerst\"{a}nde)}}
} & \parbox{74pt}{\raggedright 
\textbf{{\small Experimente (Kondensa-toren)}}
} & \parbox{83pt}{\raggedright 
\textbf{{\small Simulationen (Spannungs-quellen und Widerst\"{a}nde)}}
} & \parbox{74pt}{\raggedright 
\textbf{{\small Simulationen (Kondensa-toren)}}
} \\
\hline
\parbox{92pt}{\raggedright 
{\small Anzahl}
} & \parbox{83pt}{\raggedright 
{\small 18}
} & \parbox{74pt}{\raggedright 
{\small 10}
} & \parbox{83pt}{\raggedright 
{\small 33}
} & \parbox{74pt}{\raggedright 
{\small 7}
} \\
\hline
\parbox{92pt}{\raggedright 
{\small Erwartungswert}
} & \parbox{83pt}{\raggedright 
{\small 27,6 min}
} & \parbox{74pt}{\raggedright 
{\small 46,5 min}
} & \parbox{83pt}{\raggedright 
{\small 14,5 min}
} & \parbox{74pt}{\raggedright 
{\small 40,7 min}
} \\
\hline
\parbox{92pt}{\raggedright 
{\small Standardab-weichung}
} & \parbox{83pt}{\raggedright 
{\small 29,6 min}
} & \parbox{74pt}{\raggedright 
{\small 38,5 min}
} & \parbox{83pt}{\raggedright 
{\small 14,8 min}
} & \parbox{74pt}{\raggedright 
{\small 11,3 min}
} \\
\hline
\end{tabular}
\vspace{2pt}

}

Es fŠllt auf, dass die meisten Versuche mit Batterien und WiderstŠnde mit dem Simulationsprogramm vorgenommen worden sind. Au§erdem ist die benštigte Zeit pro Simulation deutlich geringer als beim realen Versuch. Bei Kondensatorschaltungen fŠllt dieser Unterschied jedoch nicht ins Gewicht. Da die Standardabweichungen sehr hoch sind, ist die Interpretation allerdings nicht zwingend.

\subsection{Feedback}
Eine vollstŠndige Auflistung aller Antworten findet sich in Anhang. Im Folgenden erfolgt eine Zusammenfassung der wichtigsten Antworten. Sofern das entsprechende Feedback nicht anonym abgegeben worden ist, wird der Name des SchŸlers in Klammern genannt.
Problematisch bei realen Experimenten sind GerŠte und Bauteile, die entweder defekt (Markus) oder in nicht ausreichender Anzahl vorrŠtig sind (Per, Steven, Franziska), sowie Probleme bei der Bedienung der MessgerŠte gekoppelt mit der Mšglichkeit, sie durch Fehlbedienung zu beschŠdigen (Friederike). Als Vorteil bei Computersimulationen wird angesehen, dass diese Probleme eben nicht existieren: Bauteile sind in quasi unbegrenzter Anzahl verfŸgbar, eine versehentliche BeschŠdigung ist ausgeschlossen und Probleme, die sich bei der Bedienung von MessgerŠten ergibt, treten Ÿberhaupt nicht erst auf (Alina, Franziska, Friederike, Steven). Au§erdem kšnnen Modifikationen an der Schaltung viel leichter und schneller vorgenommen werden. Andererseits werden die Probleme, die bei den realen Experimenten aufgetreten sind, auch wieder als Herausforderung begriffen. Die Vermeidung von KurzschlŸssen oder die richtige Bedienung eines Multimeters fordere wiederum eine intensivere Auseinandersetzung mit der Materie (Alina). Ebenfalls erwŠhnt worden ist, dass Experimente realer erschienen und einen direkteren Bezug zur Wirklichkeit zulie§en (Franziska) und diese letztendlich anschaulicher seien (Per). Das Simulationsprogramm dagegen liefere nur "`endlose Zahlenkolonnen"', welche nur schwer vorstellbar seien (Per). Au§erdem sei die Einarbeitungszeit relativ lang. Die Eingabe des Schaltplans sei aufwendig.
Insgesamt ist die SchŸlerschaft in zwei ungefŠhr gleich gro§e Lager gespalten, wenn nach einer PrŠferenz fŸr Experimente oder Computersimulationen gefragt wird. Auf die Frage, welche Methode das Lernen besser unterstŸtze, haben viele SchŸler geantwortet, dass man doch beides im Unterricht einsetzen solle. Einerseits seien die Simulationen gut geeignet, um den theoretischen Teil des Unterrichts zu bewerkstelligen. Andererseits wŸrden experimentelle Fertigkeiten eben nur durch tatsŠchliches Experimentieren gefšrdert werden. Eine Korrelation der PrŠferenz mit dem Leistungsstand der SchŸler ist nicht mšglich. Auch alle anderen Antworten lassen keine Aufteilung in stŠrkere und schwŠchere SchŸler zu. 
\subsection{Videoanalyse}
Von allen Doppelstunden, in denen experimentiert bzw. am Computer gearbeitet worden ist, sind Videoaufnahmen angefertigt worden. Dies hat den Vorteil, dass das Arbeitsverhalten der SchŸler genauer beobachtet werden kann, da wŠhrend der Stunde die Lehrkraft natŸrlich damit beschŠftigt gewesen ist, Hilfestellung zu leisten. Es ist jedoch zu beachten, dass die Videoaufnahmen immer nur einen Bildausschnitt zeigen, da die Optik nicht weitwinklig und der CCD-Chip nicht hochauflšsend genug gewesen sind, um den gesamten Raum abzubilden. Auch ein Schwenken der Kamera ist nur relativ selten erfolgt, da die Lehrkraft mit anderen Dingen beschŠftigt gewesen ist, was der ursprŸngliche Sinn ihres Einsatzes gewesen ist. 
Anhand der Videoaufnahmen ist ein Verlaufsprotokoll erstellt worden, von denen die hŠufigsten Ereignisse an dieser Stelle wiedergegeben werden sollen.

{\raggedright

\vspace{3pt} \noindent
\begin{tabular}{|p{369pt}|p{77pt}|}
\hline
\parbox{369pt}{\raggedright 
\textbf{{\small Ereignis}}
} & \parbox{77pt}{\raggedright 
\textbf{{\small H\"{a}ufigkeit}}
} \\
\hline
\parbox{369pt}{\raggedright 
{\small Probleme mit Laptops (sim)}
} & \parbox{77pt}{\raggedright 
{\small 2}
} \\
\hline
\parbox{369pt}{\raggedright 
{\small Defektes Messger\"{a}t / Bauteil (exp)}
} & \parbox{77pt}{\raggedright 
{\small 5}
} \\
\hline
\parbox{369pt}{\raggedright 
{\small Fehlersuche bei einer Simulation (sim)}
} & \parbox{77pt}{\raggedright 
{\small 1}
} \\
\hline
\parbox{369pt}{\raggedright 
{\small Fehlersuche bei einem experimentellen Aufbau (exp)}
} & \parbox{77pt}{\raggedright 
{\small 2}
} \\
\hline
\parbox{369pt}{\raggedright 
{\small Hilfestellung bei Computersimulation (sim)}
} & \parbox{77pt}{\raggedright 
{\small 7}
} \\
\hline
\parbox{369pt}{\raggedright 
{\small Hilfestellung bei Experiment (exp)}
} & \parbox{77pt}{\raggedright 
{\small 9}
} \\
\hline
\parbox{369pt}{\raggedright 
{\small Planlosigkeit (sim)}
} & \parbox{77pt}{\raggedright 
{\small 6}
} \\
\hline
\parbox{369pt}{\raggedright 
{\small Planlosigkeit (exp)}
} & \parbox{77pt}{\raggedright 
{\small 0}
} \\
\hline
\parbox{369pt}{\raggedright 
{\small Nicht zielgerichtetes Arbeiten (sim)}
} & \parbox{77pt}{\raggedright 
{\small 1}
} \\
\hline
\parbox{369pt}{\raggedright 
{\small Nicht zielgerichtetes Arbeiten (exp)}
} & \parbox{77pt}{\raggedright 
{\small 1}
} \\
\hline
\end{tabular}
\vspace{2pt}

}
 
Auch hier kann die statistische Relevanz bemŠngelt werden, zumal immer nur ein kleiner (jedoch zufŠllig ausgewŠhlter) Bildausschnitt betrachtet worden ist. Trotzdem ist es richtig zu sagen, dass Probleme mit defekten MessgerŠten und Bauteilen wesentlich hŠufiger aufgetreten sind als technische Probleme mit den Laptops und der darauf installierten Software. Bedarf nach sachkundiger Hilfe hat bei beiden Vorgehensweisen etwa gleich hŠufig bestanden. Erhebliche Startschwierigkeiten, bei denen SchŸler nicht gewusst haben, was sie tun mŸssen, sind ausschlie§lich bei den Computersimulationen erkennbar gewesen. Dies betrifft auch bessere SchŸler wie Alina, Friederike, Franziska und Steven, die anfangs viele Minuten nur den Bildschirm angeschaut haben. Die von der Lehrkraft erstelle Kurzanleitung, die durch ihre starke Strukturierung beim Einhalten der Arbeitsschritte den Erfolg hat garantieren sollen, hat ihren Zweck offensichtlich nicht vollstŠndig erfŸllen kšnnen und wŸrde in Zukunft einer †berarbeitung bedŸrfen (Seitens der SchŸler sind hierzu auch einige hilfreiche Anregungen gegeben worden. So wŠren zum Beispiel "`Screenshots"' in der Anleitung hilfreich, um die entsprechenden Befehle in den MenŸs einfacher zu finden). Die Lehrkraft hat daher mit den SchŸlern gemeinsam das Arbeitsblatt Schritt fŸr Schritt durchgehen mŸssen. Auch schwŠchere SchŸler wie Per und Ahmad haben danach den Arbeitsauftrag auf dem Arbeitsblatt zur Computersimulation verstanden. HŠufig suchen die SchŸler auch untereinander Rat. Hierbei lŠsst sich beobachten, dass sich schwŠchere SchŸler wie Markus Hilfe bei leistungsstŠrkeren MitschŸlern suchen, die sich mit dem gleichen Thema befassen oder befasst haben. Die AtmosphŠre ist durchwegs sehr kollegial. 
Auch die Experimente sind nicht unproblematisch gewesen. Viele Fragen (eine genaue Zahl ist aus den Videos nicht ermittelbar) hat es bezŸglich der Bedienung der MessgerŠte gegeben. Dabei ist auch stŠrkeren SchŸlern wie Daniel und Alina die Bedienung anfangs nicht leicht gefallen, da sie sich erstmal an die Unterscheidung zwischen Spannung und Strom haben erinnern mŸssen.
Nicht zielgerichtetes Arbeiten, z. B. mit den Kabeln spielen (Asgard), Verwendung des Computers zum Musikhšren (Artur) oder zum Spielen vorinstallierter Computerspiele (Quoc-Thai) ist erstaunlich selten beobachtet worden.
Weiterhin ist sehr positiv aufgefallen, dass die SchŸler kurz vor dem Ende selbststŠndig die Arbeitsmaterialien zusammengepackt und aufgerŠumt haben, ohne dass es hierzu wiederholter Aufforderungen bedurft hŠtte. Alle SchŸler haben sich am gemeinsamen AufrŠumen beteiligt.
In den Einzelstunden (von denen es jedoch keine Videoaufzeichnungen gibt) ist aufgefallen, dass die mŸndliche Beteiligung gegenŸber dem ersten Halbjahr deutlich gestiegen ist. 

\section{Interpretation}
Probleme mit der geringen Stichprobengrš§e au§er Acht gelassen, ist das sich ergebene Bild weder mit den hier aufgestellten Hypothesen, noch mit den Aussagen von Finkelstein und Mitarbeiter inkonsistent. Am effektivsten erweisen sich die Simulationen beim Themenkomplex "Spannungsquellen"'. Da Quiz \#2 jedoch eine exakte Kopie von Quiz \#1 gewesen ist, kann hier nicht auf eine Verbesserung im konzeptionellen VerstŠndnis geschlossen werden. Dies sieht anders beim nŠchsten Themenkomplex "Widerstandsschaltungen"' aus. Hier sind die Aufgaben beim Abschlusstest verŠndert worden, sodass ein konzeptionelles VerstŠndnis fŸr die Aufgabenart erforderlich ist, weshalb es nicht unrealistisch ist anzunehmen, dass sich unter Verwendung von Computersimulationen tatsŠchlich Vorteile fŸr das konzeptionelle VerstŠndnis ergeben. Bei den Kondensatorschaltungen ist zwischen Experiment und Simulation kein nennenswerter Vorteil erkennbar. Betrachtet man nur die Anzahl der SchŸler, die sich nach Fertigstellung des praktischen Teils verbessert haben, so ist kein Unterschied erkennbar. AuffŠllig ist jedoch die gro§e Zahl an SchŸlern, die sich nach dem Aufbau der realen Schaltungen verschlechtert haben. GesprŠche mit einigen SchŸlern haben zu der Erkenntnis gefŸhrt, dass sich Fehlinterpretationen der von den MessgerŠten gelieferten Anzeigeergebnisse verwirrend und damit auf die Leistung negativ ausgewirkt haben. Wenn dem tatsŠchlich so sein sollte, so kšnnte ein selbstgesteuertes Experimentieren ohne ausfŸhrliche RŸckmeldung zu falschen oder verfŠlschenden Ergebnissen fŸhren.
Die Zeitersparnis pro Simulation fŠllt bei den Kondensatorschaltungen wesentlich geringer aus als bei den Schaltungen mit WiderstŠnden und Spannungsquellen. Mšglicherweise deutet dies auf einen relativen hohen Zeitaufwand hin, der zum Einarbeiten notwendig ist. Hinweise hierzu finden sich auf den Feedback-Bšgen, die jedoch nicht explizit zwischen den Kondensatorschaltungen und den anderen Schaltungen unterscheiden. Dies kšnnte darauf hindeuten, dass das Material zur Vorbereitung bei den Kondensatorschaltungen missverstŠndlich oder nicht ausfŸhrlich genug gewesen ist. Insgesamt betrachtet kann jedoch festgestellt werden, dass Computersimulationen weniger zeitaufwendig sind und die SchŸler somit im gleichen Zeitraum mehr Simulationen als Experimente durchfŸhren haben kšnnen. Es existiert eine Korrelation zwischen der hohen Zahl an durchgefŸhrten Simulationen und dem besseren Abschneiden bei den Quiz. Wie bei jeder Statistik muss aber auch hier darauf hingewiesen werden, dass eine blo§e Korrelation nicht zwangslŠufig  einen Kausalzusammenhang nach sich zieht. Trotzdem liegt die Vermutung nahe, dass durch die Entlastung der SchŸler von Routinearbeiten und die BeschŠftigung mit idealisierten und funktionstŸchtigen Bauteilen dazu fŸhrt, dass die SchŸler nicht durch StšreinflŸsse von den grundlegenden Konzepten abgelenkt werden. 
Wie aus den Videoaufnahmen und den Feedback-Bšgen der SchŸler hervorgeht, ist ein solcher "Stšreinfluss"' die Bedienung der MessgerŠte gewesen. Offensichtlich hat es hier ganz erhebliche Schwierigkeiten gegeben, weshalb bereits wŠhrend der Untersuchung entsprechende Grundlagen thematisiert worden sind. Die ursprŸngliche Annahme, dass die SchŸler bereits Ÿber genug experimentelle Fertigkeiten verfŸgen und diese daher nicht mehr explizit trainiert werden mŸssen (weshalb einem Einsatz von Programmen wie PSPICE nichts im Weg steht), hat sich damit als falsch herausgestellt. Finkelstein et al. haben festgestellt, dass Computersimulationen auch das experimentelle Kšnnen fšrdern. So haben Studenten, die sich vorher nur mit den Simulationen beschŠftigt haben, bei einem darauffolgenden experimentellen Teil wesentlich besser abgeschnitten. Dies ist jedoch nicht weiter verwunderlich, weil das dort eingesetzte Simulationsprogramm die Bedienung eines Multimeters gleich mitsimuliert hat. Insofern liegt kein Widerspruch vor. 
Interessanterweise sind sich die SchŸler Ð wie aus den Feedback-Bšgen hervorgeht Ð der Problematik bewusst. Sie sehen einerseits klare Vorteile von PSPICE, wie die potentielle Zeitersparnis und das Vermeiden von Routinearbeiten, sind sich andererseits aber durchaus im klaren, dass sie diese Routinearbeiten noch nicht sicher beherrschen. Zu diesen Routinearbeiten gehšrt auch die Ð nicht vermeidbare Ð Fehlersuche, die sowohl den SchŸlern als auch der Lehrkraft einen hohen Zeitaufwand abverlangt. Auch die systematische Fehlersuche ist eine wichtige Fertigkeit, die jedoch Ð wenn sie einmal beherrscht worden ist, nur weiteren lŠstigen Zeitaufwand nach sich zieht. FŸr den Kurs ist es sicherlich produktiver, wenn sich die Lehrkraft mit den konzeptionellen Problemen der SchŸler beschŠftigen kann als sich in dicken KabelbŠumen zurechtfinden muss. Im Šu§ersten Notfall kann den SchŸlern eine Kopie einer lauffŠhigen Simulationsdatei gegeben werden, die sie dann nur noch verŠndern brauchen. 
Zum Erwerb dieser Kompetenzen sind Computersimulationen dennoch nicht zielfŸhrend, da sie das Problem einfach umgehen. Auch dies ist in den Feedback-Bšgen relativ direkt zur Sprache gekommen. Dass die Bedienung eines Multimeters den SchŸlern noch vor extrem gro§e HŸrden gestellt hat, ist ebenfalls in der Videoanalyse sichtbar geworden. Da bereits wŠhrend der DurchfŸhrung des Unterrichtsversuchs hier Defizite erkennbar gewesen sind, hat die Lehrkraft einige der Einzelstunden der Messung von Spannungen und Stršmen sowie der Bedeutung dieser Grš§en gewidmet. PSPICE kann eingesetzt werden, um die mit den Grundlagen vertrauten SchŸlern gezielt zu fšrdern, auch wenn eine PrŠferenz fŸr eine Methode keiner bestimmten SchŸlergruppe zugeordnet werden kann.
Es scheint daher so, als wenn Computersimulationen FŠhigkeiten fšrdern, die bevorzugt in schriftlicher Form abgefragt werden kšnnen. Es muss jedoch positiv erwŠhnt werden, dass der Erfolg der Computersimulationen nicht allein auf Umgehung nicht abfragbarer Schwierigkeiten gegrŸndet ist. DafŸr spricht, dass sich die SchŸler wesentlich intensiver mit Schaltungen auseinandersetzen haben kšnnen, da sie in der gleichen Zeit mehr Ergebnisse erhalten haben. Offensichtlich gibt es also weitere Schwierigkeiten, die bei der Verwendung von Computersimulationen nicht auftreten. So sind Computersimulationen im Gegensatz zu den unŸbersichtlichen KabelbŠumen, die bei den realen Experimenten aufgetreten sind, relativ Ÿbersichtlich (aus Feedback). Auch der Lehrkraft hat es viel Zeit gekostet, sich bei der Fehlersuche durch die Verkabelung zu arbeiten (aus Videoanalyse). Trotzdem sollte es mšglich sein, eine einmal funktionierend aufgebaute Schaltung schnell zu verŠndern, z. B. durch den Wechsel eines Widerstands mit einem anderen Wert. Nach den €u§erungen auf den Feedback-Bšgen ist die Verwendung eines Computers vorteilhaft, da so Variationen einer Schaltung wesentlich schneller haben vorgenommen werden kšnnen. Mšglicherweise mŸssen an dieser Stelle fehlende und defekte Arbeitsmaterialien genannt werden, welche die Arbeit erschwert haben (aus Feedback). Selbst WiderstŠnde sind nur in relativ begrenzter Anzahl verfŸgbar gewesen. Am problematischsten hierbei sind aber vermutlich die Multimeter gewesen, von denen sich einige SchŸler welche haben teilen mŸssen. Bei vielen Experimenten sind auch mehrere Strom- und Spannungsmessungen sinnvoll. Dazu muss das Multimeter neu angeschlossen und der Messbereich entsprechend eingestellt werden. Dieser Arbeitsaufwand entfŠllt bei den Simulationen, da das Programm automatisch alle Stršme und Spannungen ausgibt, also auch solche, fŸr die sich SchŸler auf den ersten Blick vielleicht nicht interessieren wŸrden. 

Computer sind Ð wie aus der Videoanalyse ersichtlich ist Ð nicht immun vor Problemen. Sie lassen sich aber von der Lehrkraft in der Regel schneller lšsen und sind hŠufig schon vor Unterrichtsbeginn erkennbar. Dies entspricht weitestgehend den Erfahrungen von Finkelstein und Mitarbeitern. Allerdings ist der notwendige Zeitaufwand, um sich in die Bedienung von PSPICE einzuarbeiten, nicht unerheblich. Dies gilt auch fŸr gute SchŸler wie Alina, Steven, Friederike und Franziska. Allgemeine Computerkenntnisse wie sie zum Starten eines Programms benštigt werden, sind natŸrlich erforderlich, allerdings in der heutigen Zeit in den meisten FŠllen vorhanden. Bei kleineren Problemen (wie sie z. B. beim Verschieben von Dateien vorkommen kšnnen) haben sich die SchŸler untereinander geholfen, ohne dass eine Intervention der Lehrkraft nštig gewesen ist. Es gibt keine Anzeichen dafŸr, dass der Computer selbst ein HŸrde dargestellt hŠtte. Trotzdem dauert es natŸrlich seine gewisse Zeit, bis die SchŸler alle notwendigen Dateien auf den DatentrŠgern lokalisiert haben. Dass eine einmalige Einarbeitung notwendig ist, ist natŸrlich nicht Ÿberraschend und bei der Planung in Kauf genommen worden. Dazu hat auch beigetragen, dass die gewŠhlte textbasierte Betriebsart von PSPICE eben nicht als všllig intuitiv und visuell bezeichnet werden kann. Insofern kann den Aussagen der SchŸler zugestimmt werden, dass mit dem Aufbau einer realen Schaltungen sofort begonnen werden kann. Als weiterer Vorteil ist genannt worden, dass eine Verbindung zwischen den Experimenten und PhŠnomenen des tŠglichen Lebens (z. B. Stromversorgung) einfacher zu erkennen gewesen ist als mit Simulationen. Seitens der SchŸlerschaft wird daher zu Recht ein komplementŠrer Einsatz sowohl von Simulationen als auch von realen Experimenten vorgeschlagen. Von einer generellen Abneigung gegenŸber PSPICE kann somit keine Rede sein. Auf die Frage, welche Methode letztendlich bevorzugt wird, sind Experimente und Simulationen zu fast gleichen Teilen angegeben worden. Einige SchŸler haben sogar angefragt, ob sie mit dem Programm zu Hause weiterarbeiten kšnnten (um die Vergleichbarkeit sicherzustellen, ist die Internetadresse der Webseite, von der das Programm installiert werden kann, erst nach dem Ende des Unterrichtsversuchs bekanntgegeben worden).
Die bei den Versuchen und Simulationen aufgetretenen Probleme sind Anlass fŸr viele (englische) GesprŠche der SchŸler untereinander gewesen. Dies zeugt von einem hohen Grad an Kompetenz im Bereich der Kommunikation, da das Problem bzw. die erzielten Ergebnisse genau dargestellt werden mŸssen.

\section{Alternativen}
Ein Hauptkritikpunkt, der gegen den Einsatz von PSPICE spricht, ist die geringe Anschaulichkeit, sowohl der Eingabe- als auch der Ausgabedateien. Eine VerknŸpfung dieser ASCII-Dateien mit Alltagsproblemen ist den SchŸlern nicht immer sofort offensichtlich gewesen (aus Feedback). Dieses Problem ist bewusst in Kauf genommen worden, da es bei der Untersuchung darum geht, den Einfluss der starken Veranschaulichung, von der in den amerikanischen Simulationsprogrammen Gebrauch gemacht wird, auszuschlie§en. Der Autor ist der Ansicht, dass der Einsatz des Programms trotzdem nicht zu einem Mangel an AuthentizitŠt \cite{Muckenfuss:2006wb} des Unterrichts fŸhrt, weil es sich bei PSPICE um ein Programm handelt, welches in der Industrie tatsŠchlich eingesetzt wird und sich dort bewŠhrt hat. Auf diesen Umstand ist im Unterricht wiederholt hingewiesen worden. Trotzdem sind Zweifel berechtigt. Auch wenn PSPICE fŸr die zu untersuchende Fragestellung geeignet gewesen ist, so stellt sich dessen Einsatz im Unterricht nicht ganz problemlos dar. Die drei offensichtlichen Alternativen lauten entweder ganz auf den Einsatz von Simulationen zugunsten von realen Experimenten zu verzichten, PSPICE im interaktiven Graphikmodus zu betreiben oder die vom PhET entwickelten Anwendungen zu verwenden, die wesentlich visueller sind und bei denen die benštigten Hilfsmittel Teil der Simulation sind. Einem Einsatz von PSPICE steht prinzipiell nichts im Wege. Er lohnt sich jedoch nur bei langen Experimentierserien, bei denen die Zeitersparnis sich als vorteilhaft herausstellt. Au§erdem sollten gewisse experimentelle Fertig- und FŠhigkeiten bereits vorhanden sein, da PSPICE nichts zu deren Erwerb beitragen kann. Diese Voraussetzung ist wider Erwarten nicht erfŸllt gewesen. Speziell die Unterscheidung zwischen Spannung und Strom hat den SchŸlern noch erhebliche Schwierigkeiten bereitet, die sich dann bei deren Messung fortgesetzt haben. In den Einzelstunden sind diese Aspekte daher immer thematisiert worden. Idealerweise wŠre dies aber bereits vor Beginn der Experimentier- / Simulationsphase geschehen. Zu Beginn des ersten Halbjahres haben die SchŸler einen Kompetenztest geschrieben, mit dessen Hilfe die Vorkenntnisse der SchŸler genau ermittelt werden sollten. Der Schwerpunkt hat jedoch im Bereich der Mechanik gelegen. In den darauffolgenden zwei Wochen sind alle Themengebiete kurz wiederholt worden, die fŸr die beiden Halbjahre von Relevanz gewesen sind. Besonders berŸcksichtigt worden sind allerdings die Themenbereiche aus dem Kompetenztest, mit denen die SchŸler besondere Schwierigkeiten gehabt haben.
Entsprechende Vorkenntnisse hŠtten womšglich auch dazu gefŸhrt, dass die Ein- und Ausgabedateien weniger abstrakt gewirkt hŠtten, da eine VerknŸpfung mit bereits Bekanntem einfacher mšglich gewesen wŠre. Ob die blo§e Zuschaltung des grafischen Schaltplaneditors in PSPICE ausgereicht hŠtte, um eine stŠrke Anschaulichkeit zu erzielen, ist nicht sicher, da die SchŸler durchaus mit SchaltplŠnen jeder Schaltung gearbeitet haben, die sie jedoch selber haben zeichnen mŸssen. Die interaktive Betriebsart hat vielmehr den Nachteil, dass die gro§e Bauteilebibliothek und der Befehlsumfang dazu gefŸhrt hŠtten, dass die SchŸler ihrer natŸrlichen Neugierde folgend mit viel zu komplexen Schaltungen gespielt hŠtten. Finkelstein und Mitarbeiter schreiben darŸber Ÿber ihre eigenen Programme : ``As discussed below, computers can [...] constrain the students in productive ways.'' Dabei sei erwŠhnt, dass die SchŸler meistens produktiv gearbeitet haben und nicht mit den Materialien sachfremd gespielt haben (aus Videoanalyse). Auch (oder gerade) Computer lassen sich fŸr Zwecke einsetzen, die dem gesetzten Ziel nicht fšrderlich sind ("`Solitaire"' oder andere Spiele). Trotzdem erscheint es richtig, die SchŸler nicht mit nichtbenštigtem Material bzw. anderen Freiheitsgeraden zu konfrontieren, die von der eigentlichen Aufgabenstellung nur ablenken. Damit ist PSPICE auch im interaktiven Grafikbetrieb nicht die Ideallšsung. Die beste Wahl fŸr den betrachteten Kurs wŠre damit vermutlich das von der PhET entwickelte circuit construction lab gewesen, welches die Vorteile der einfachen VerŠnderbarkeit einer Schaltung mit der Veranschaulichung durch entsprechende Visualisierung und dem Mitsimulieren der benštigten MessgerŠte kombiniert. Allerdings unterstŸtzt letzteres weder Kondensator- noch Wechselstromschaltungen, weshalb es sich nur zum Erlernen der Grundprinzipien anhand von Widerstandsschaltungen eignet. Zum VerstŠndnis des Kondensators wŠre somit doch wieder ein Programm wie PSPICE notwendig.
Ein weiterer Vorteil von Simulationen soll nicht verschwiegen werden: In Zeiten finanzieller Haushaltsnšte ist die Ausstattung der PhysikrŠume nicht immer auf den Stand, den man sich wŸnschen wŸrde. 

\section{Fazit und Ausblick}
Der Autor sieht den Ausgang des Unterrichtsversuchs mit gemischten GefŸhlen, weil er Ÿber die Studienergebnisse aus Colorado zu enthusiastisch gewesen ist. Dies ist auf den ersten Blick erstaunlich, da ein wirklicher Widerspruch zu den amerikanischen Ergebnissen nicht erkennbar ist. WidersprŸchlich dagegen sind die gestellten Anforderungen. Einerseits soll durch die Eliminierung von Routinearbeiten die Motivation erhšht werden. Andererseits ist die Beherrschung jener Routineaufgaben wichtig und kann nur durch Routine erlernt werden. 
In der Tat mšgen Computersimulationen letztendlich nŸtzlicher sein, wenn Erfolg allein mit Hilfe des Abschneidens in Klausuren und Tests gemessen wird. Dies legen nicht nur die Ergebnisse der Tests nahe, sondern ist auch von den SchŸlern selber bemerkt worden. Die Aussagen, dass am besten beide Methoden zum Einsatz kommen sollen, da sie verschiedene FŠhigkeiten fšrdern, deutet darauf hin, dass beides komplementŠr im Unterricht nebenher verwendet werden sollte. Im Umkehrschluss bedeutet dies, dass die Computersimulationen reale Experimente nicht ersetzen kšnnen, da sie zur Auseinandersetzung mit Problemen fŸhren, die bei Computersimulationen einfach nicht auftauchen. So ist es sicherlich wŸnschenswert, wenn ein Abiturient problemlos ein MessgerŠt bedienen oder systematisch nach Fehlern suchen kann. Bei einer ausschlie§lich auf Klausur- und Testergebnissen basierender RŸckmeldung wŠre dieses Problem mšglicherweise nie offensichtlich geworden. Experimentelle Klausuren (bzw. Tests) haben sowohl in der Schule als auch an der UniversitŠt Seltenheitswert. 
Abschlie§end betrachtet kann gesagt werden, dass ein offener Unterricht wie er hier zum Einsatz gekommen ist, viele Vorteile aufweist. Der Erfolg setzt aber eine umfangreiche RŸckmeldung seitens der Lerngruppe voraus, damit die Lehrkraft auch tatsŠchlich beratend tŠtig werden kann.

\pagebreak

\bibliography{bibsimexp}

\appendix

\pagebreak

\section{Auswertung Feedback}

\begin{indentation}{0pt}{0pt}{0pt}
{\small Im Folgenden werden alle Antworten aufgelistet, die von den
Sch\"{u}lern gegeben worden sind. Mehrfachnennungen werden mit einer
entsprechenden Angabe der H\"{a}ufigkeit in Klammern angegeben.}
\end{indentation}

\begin{indentation}{0pt}{0pt}{0pt}

\end{indentation}

\subsection{Probleme Experiment}

\begin{indentation}{0pt}{0pt}{0pt}
{\small Defekte Dr\"{a}hte}
\end{indentation}

\begin{indentation}{0pt}{0pt}{0pt}
{\small Defekte Messger\"{a}te}
\end{indentation}

\begin{indentation}{0pt}{0pt}{0pt}
{\small Defekte Batterien}
\end{indentation}

\begin{indentation}{0pt}{0pt}{0pt}
{\small Kabelsalat}
\end{indentation}

\begin{indentation}{0pt}{0pt}{0pt}
{\small Probleme mit der Bedienung der Messger\"{a}te (x3)}
\end{indentation}

\begin{indentation}{0pt}{0pt}{0pt}
{\small Begrenzte M\"{o}glichkeiten [da nicht alle Teile beliebig
verf\"{u}gbar]}
\end{indentation}

\begin{indentation}{0pt}{0pt}{0pt}
{\small Fehler schwer einzugrenzen (x2)}
\end{indentation}

\begin{indentation}{0pt}{0pt}{0pt}
{\small Unabsichtliche Besch\"{a}digung eines Bauteils}
\end{indentation}

\begin{indentation}{0pt}{0pt}{0pt}

\end{indentation}

\subsection{Probleme Simulation}

\begin{indentation}{0pt}{0pt}{0pt}
{\small Endlose Zahlenkolonnen}
\end{indentation}

\begin{indentation}{0pt}{0pt}{0pt}
{\small Schwer bildlich vorstellbar (x2)}
\end{indentation}

\begin{indentation}{0pt}{0pt}{0pt}
{\small Einarbeitungszeit erforderlich (x4)}
\end{indentation}

\begin{indentation}{0pt}{0pt}{0pt}
{\small \"{U}bersichtlicher}
\end{indentation}

\begin{indentation}{0pt}{0pt}{0pt}
{\small Keine Grafik bei der Schaltplaneingabe}
\end{indentation}

\begin{indentation}{0pt}{0pt}{0pt}
{\small Manchmal keine Ver\"{a}nderung sichtbar [durch falsche
Ver\"{a}nderung von Parametern]}
\end{indentation}

\begin{indentation}{0pt}{0pt}{0pt}

\end{indentation}

\subsection{Vorteile Experiment}

\begin{indentation}{0pt}{0pt}{0pt}
{\small Einfacher zu "`sehen"' was man tut (x2)}
\end{indentation}

\begin{indentation}{0pt}{0pt}{0pt}
{\small Macht Spa\ss{}}
\end{indentation}

\begin{indentation}{0pt}{0pt}{0pt}
{\small Anschaulich (x4)}
\end{indentation}

\begin{indentation}{0pt}{0pt}{0pt}
{\small Einfachere \"{U}bertragung auf reale Ph\"{a}nomene, Stromnetz
etc.}
\end{indentation}

\begin{indentation}{0pt}{0pt}{0pt}
{\small Mehr Freiraum }
\end{indentation}

\begin{indentation}{0pt}{0pt}{0pt}
{\small Funktionsgraphen werden vom Programm automatisch erstellt
(x2)}
\end{indentation}

\begin{indentation}{0pt}{0pt}{0pt}
{\small Mehr Nachdenken erforderlich um Fehler zu vermeiden
(Kurzschl\"{u}sse etc.) (x2)}
\end{indentation}

\begin{indentation}{0pt}{0pt}{0pt}
{\small Erscheinen realer}
\end{indentation}

\begin{indentation}{0pt}{0pt}{0pt}
{\small Auch komplizierte Schaltungen k\"{o}nnen schnell aufgebaut
werden}
\end{indentation}

\begin{indentation}{0pt}{0pt}{0pt}

\end{indentation}

\subsection{Vorteile Simulation}

\begin{indentation}{0pt}{0pt}{0pt}
{\small Fehler einfacher zu entdecken}
\end{indentation}

\begin{indentation}{0pt}{0pt}{0pt}
{\small Keine Bedienungsfehler an den Messger\"{a}ten (x2)}
\end{indentation}

\begin{indentation}{0pt}{0pt}{0pt}
{\small Wesentlich leichter die Schaltung schnell zu modifizieren
(x4)}
\end{indentation}

\begin{indentation}{0pt}{0pt}{0pt}
{\small Enorme Zeitersparnis (x4)}
\end{indentation}

\begin{indentation}{0pt}{0pt}{0pt}
{\small Mehr Freiraum [Teile beliebig verf\"{u}gbar] (x2)}
\end{indentation}

\begin{indentation}{0pt}{0pt}{0pt}

\end{indentation}

\subsection{Was ist besser f\"{u}r das Lernen geeignet?}

\begin{indentation}{0pt}{0pt}{0pt}
{\small Simulationen nachdem man mit realen Experimenten vertraut ist
(x2)}
\end{indentation}

\begin{indentation}{0pt}{0pt}{0pt}
{\small Beide zusammen (x2)}
\end{indentation}

\begin{indentation}{0pt}{0pt}{0pt}
{\small Simulationen durch die Vereinfachung des Prozesses}
\end{indentation}

\begin{indentation}{0pt}{0pt}{0pt}
{\small Simulationen da ,"`Messwerte"' verl\"{a}sslicher}
\end{indentation}

\begin{indentation}{0pt}{0pt}{0pt}
{\small Experimente wegen ihres spielerischen Charakters}
\end{indentation}

\begin{indentation}{0pt}{0pt}{0pt}
{\small Experimente besser da mehr Vor\"{u}berlegungen notwendig (x3)}
\end{indentation}

\begin{indentation}{0pt}{0pt}{0pt}
{\small Experimente besser, da der Computer die ganze Arbeit macht}
\end{indentation}

\begin{indentation}{0pt}{0pt}{0pt}
{\small Simulationen f\"{u}r die Theorie, Experimente f\"{u}r die
Anwendungen}
\end{indentation}

\begin{indentation}{0pt}{0pt}{0pt}

\end{indentation}

\begin{indentation}{0pt}{0pt}{0pt}

\end{indentation}

\subsection{Was m\"{o}gt ihr lieber?}

\begin{indentation}{0pt}{0pt}{0pt}
{\small Simulationen (x4)}
\end{indentation}

\begin{indentation}{0pt}{0pt}{0pt}
{\small Experimente (x5)}
\end{indentation}

\begin{indentation}{0pt}{0pt}{0pt}

\end{indentation}

\section{Auswertung Videoanalyse}

{\raggedright

\vspace{3pt} \noindent
\begin{tabular}{p{33pt}p{400pt}}
\parbox{33pt}{\raggedright 
\textbf{\textsf{{\small 7. Mai}}}
} & \parbox{400pt}{\raggedright 

} \\
\parbox{33pt}{\raggedright 

} & \parbox{400pt}{\raggedright 

} \\
\parbox{33pt}{\raggedright 
\textsf{{\small sim}}
} & \parbox{400pt}{\raggedright 
\textsf{{\small Steven, Janna, Markus}}
} \\
\parbox{33pt}{\raggedright 
\textsf{{\small exp}}
} & \parbox{400pt}{\raggedright 
\textsf{{\small Ahmad, Pascal, Daniel, Asgard, Alina}}
} \\
\parbox{33pt}{\raggedright 
\textsf{{\small exp/sim}}
} & \parbox{400pt}{\raggedright 
\textsf{{\small Quoc-Thai, Artur}}
} \\
\parbox{33pt}{\raggedright 

} & \parbox{400pt}{\raggedright 

} \\
\parbox{33pt}{\raggedright 
\textsf{{\small 2 min}}
} & \parbox{400pt}{\raggedright 
\textsf{{\small L holt mehr Kabel}}
} \\
\parbox{33pt}{\raggedright 
\textsf{{\small 4 min}}
} & \parbox{400pt}{\raggedright 
\textsf{{\small L hilft am Computer}}
} \\
\parbox{33pt}{\raggedright 
\textsf{{\small 7 min}}
} & \parbox{400pt}{\raggedright 
\textsf{{\small S arbeiten zielgerichtet}}
} \\
\parbox{33pt}{\raggedright 
\textsf{{\small 12 min}}
} & \parbox{400pt}{\raggedright 
\textsf{{\small L besch\"{a}ftigt mit defektem Me\ss{}ger\"{a}t}}
} \\
\parbox{33pt}{\raggedright 
\textsf{{\small 15 min}}
} & \parbox{400pt}{\raggedright 
\textsf{{\small L betrachtet bei der Fehlersuche verschachtelte
Kabelb\"{a}ume}}
} \\
\parbox{33pt}{\raggedright 
\textsf{{\small 18 min}}
} & \parbox{400pt}{\raggedright 
\textsf{{\small Markus sucht Hilfe bei Mitsch\"{u}lern}}
} \\
\parbox{33pt}{\raggedright 
\textsf{{\small 26 min}}
} & \parbox{400pt}{\raggedright 
\textsf{{\small Artur und Quoc-Thai suchen Hilfe}}
} \\
\parbox{33pt}{\raggedright 
\textsf{{\small 28 min}}
} & \parbox{400pt}{\raggedright 
\textsf{{\small L hilft Alina}}
} \\
\parbox{33pt}{\raggedright 
\textsf{{\small 32 min}}
} & \parbox{400pt}{\raggedright 
\textsf{{\small L ersetzt ein Me\ss{}ger\"{a}t}}
} \\
\parbox{33pt}{\raggedright 
\textsf{{\small 33 min}}
} & \parbox{400pt}{\raggedright 
\textsf{{\small L ersetzt ein defektes Kabel}}
} \\
\parbox{33pt}{\raggedright 
\textsf{{\small 34 min}}
} & \parbox{400pt}{\raggedright 
\textsf{{\small L ersetzt ein weiteres Me\ss{}ger\"{a}t}}
} \\
\parbox{33pt}{\raggedright 
\textsf{{\small 35 min}}
} & \parbox{400pt}{\raggedright 
\textsf{{\small Steven hat eine Frage}}
} \\
\parbox{33pt}{\raggedright 
\textsf{{\small 39 min}}
} & \parbox{400pt}{\raggedright 
\textsf{{\small L hilft Steven}}
} \\
\parbox{33pt}{\raggedright 
\textsf{{\small 40 min}}
} & \parbox{400pt}{\raggedright 
\textsf{{\small Asgard verwendet Kabel f\"{u}r nicht zielf\"{u}hrende
Zwecke}}
} \\
\parbox{33pt}{\raggedright 
\textsf{{\small 44 min}}
} & \parbox{400pt}{\raggedright 
\textsf{{\small L hilft bei Messung}}
} \\
\parbox{33pt}{\raggedright 
\textsf{{\small 47 min}}
} & \parbox{400pt}{\raggedright 
\textsf{{\small wieder Probleme mit einem Me\ss{}ger\"{a}t}}
} \\
\parbox{33pt}{\raggedright 
\textsf{{\small 50 min}}
} & \parbox{400pt}{\raggedright 
\textsf{{\small Asgard sieht unbeteiligt aus}}
} \\
\parbox{33pt}{\raggedright 
\textsf{{\small 57 min}}
} & \parbox{400pt}{\raggedright 
\textsf{{\small Artur h\"{o}rt Musik mit dem Laptop (\"{u}ber
Kopfh\"{o}rer)}}
} \\
\parbox{33pt}{\raggedright 
\textsf{{\small 58 min}}
} & \parbox{400pt}{\raggedright 
\textsf{{\small L hilft Asgard und Daniel}}
} \\
\parbox{33pt}{\raggedright 
\textsf{{\small 59 min}}
} & \parbox{400pt}{\raggedright 
\textsf{{\small L hilft Alina und Ahmad}}
} \\
\parbox{33pt}{\raggedright 
\textsf{{\small 60 min}}
} & \parbox{400pt}{\raggedright 
\textsf{{\small S r\"{a}umen auf}}
} \\
\parbox{33pt}{\raggedright 

} & \parbox{400pt}{\raggedright 

} \\
\parbox{33pt}{\raggedright 

} & \parbox{400pt}{\raggedright 

} \\
\parbox{33pt}{\raggedright 
\textbf{\textsf{{\small 28. Mai}}}
} & \parbox{400pt}{\raggedright 

} \\
\parbox{33pt}{\raggedright 

} & \parbox{400pt}{\raggedright 

} \\
\parbox{33pt}{\raggedright 
\textsf{{\small sim}}
} & \parbox{400pt}{\raggedright 
\textsf{{\small Asgard, Quoc-Thai, Artur, Per, Ahmad, Markus, Janna}}
} \\
\parbox{33pt}{\raggedright 
\textsf{{\small exp}}
} & \parbox{400pt}{\raggedright 
\textsf{{\small Alina, Friederike, Franziska, Daniel, Steven}}
} \\
\parbox{33pt}{\raggedright 

} & \parbox{400pt}{\raggedright 

} \\
\parbox{33pt}{\raggedright 

} & \parbox{400pt}{\raggedright 
\textsf{{\small L hilft Alina, Friederike und Franziska}}
} \\
\parbox{33pt}{\raggedright 

} & \parbox{400pt}{\raggedright 

} \\
\parbox{33pt}{\raggedright 
\textbf{\textsf{{\small 4. Juni}}}
} & \parbox{400pt}{\raggedright 

} \\
\parbox{33pt}{\raggedright 

} & \parbox{400pt}{\raggedright 

} \\
\parbox{33pt}{\raggedright 
\textsf{{\small sim}}
} & \parbox{400pt}{\raggedright 
\textsf{{\small Ahmad, Per, Asgard, Janna}}
} \\
\parbox{33pt}{\raggedright 
\textsf{{\small exp}}
} & \parbox{400pt}{\raggedright 
\textsf{{\small Markus, Daniel, Steven}}
} \\
\parbox{33pt}{\raggedright 

} & \parbox{400pt}{\raggedright 

} \\
\parbox{33pt}{\raggedright 
\textsf{{\small 8 min}}
} & \parbox{400pt}{\raggedright 
\textsf{{\small Ahmad und Per haben Probleme mit dem Verst\"{a}ndnis
des Arbeitsauftrags}}
} \\
\parbox{33pt}{\raggedright 
\textsf{{\small 20 min}}
} & \parbox{400pt}{\raggedright 
\textsf{{\small L sucht Fehler in Schaltung (exp)}}
} \\
\parbox{33pt}{\raggedright 
\textsf{{\small 28 min}}
} & \parbox{400pt}{\raggedright 
\textsf{{\small Fehlersuche abgeschlossen}}
} \\
\parbox{33pt}{\raggedright 
\textsf{{\small 28 min}}
} & \parbox{400pt}{\raggedright 
\textsf{{\small Fehlersuche bei einer Simulation}}
} \\
\parbox{33pt}{\raggedright 
\textsf{{\small 30 min}}
} & \parbox{400pt}{\raggedright 
\textsf{{\small Fehlersuche abgeschlossen}}
} \\
\parbox{33pt}{\raggedright 

} & \parbox{400pt}{\raggedright 

} \\
\parbox{33pt}{\raggedright 

} & \parbox{400pt}{\raggedright 

} \\  
\end{tabular}
}
{\raggedright

\vspace{3pt} \noindent
\begin{tabular}{p{33pt}p{400pt}}

\parbox{33pt}{\raggedright 
\textbf{\textsf{{\small 11. Juni}}}
} & \parbox{400pt}{\raggedright 

} \\
\parbox{33pt}{\raggedright 

} & \parbox{400pt}{\raggedright 

} \\
\parbox{33pt}{\raggedright 
\textsf{{\small sim}}
} & \parbox{400pt}{\raggedright 
\textsf{{\small Markus, Pascal, Franziska, Friederike, Alina, Steven,
Daniel}}
} \\
\parbox{33pt}{\raggedright 
\textsf{{\small exp}}
} & \parbox{400pt}{\raggedright 
\textsf{{\small Janna, Artur, Quoc-Thai}}
} \\
\parbox{33pt}{\raggedright 

} & \parbox{400pt}{\raggedright 

} \\
\parbox{33pt}{\raggedright 
\textsf{{\small 9 min}}
} & \parbox{400pt}{\raggedright 
\textsf{{\small L muss Laptop austauschen}}
} \\
\parbox{33pt}{\raggedright 
\textsf{{\small 16 min}}
} & \parbox{400pt}{\raggedright 
\textsf{{\small L immer noch mit Laptop besch\"{a}ftigt}}
} \\
\parbox{33pt}{\raggedright 
\textsf{{\small 20 min}}
} & \parbox{400pt}{\raggedright 
\textsf{{\small Artur ben\"{o}tigt ein Ersatznetzteil}}
} \\
\parbox{33pt}{\raggedright 
\textsf{{\small 21 min}}
} & \parbox{400pt}{\raggedright 
\textsf{{\small Friederike, Alina, Franziska und Steven starren auf
die Bildschirme}}
} \\
\parbox{33pt}{\raggedright 
\textsf{{\small 22 min}}
} & \parbox{400pt}{\raggedright 
\textsf{{\small Probleme mit einem anderen Laptop}}
} \\
\parbox{33pt}{\raggedright 
\textsf{{\small 23 min}}
} & \parbox{400pt}{\raggedright 
\textsf{{\small L geht mit Friederike, Alina und Franziska das
Arbeitsblatt gemeinsam durch}}
} \\
\parbox{33pt}{\raggedright 
\textsf{{\small 34 min}}
} & \parbox{400pt}{\raggedright 
\textsf{{\small S haben das Arbeitsblatt verstanden}}
} \\
\parbox{33pt}{\raggedright 
\textsf{{\small 35 min}}
} & \parbox{400pt}{\raggedright 
\textsf{{\small L hilft Steven mit der Simulation}}
} \\
\parbox{33pt}{\raggedright 
\textsf{{\small 37 min}}
} & \parbox{400pt}{\raggedright 
\textsf{{\small S arbeiten alle selbstst\"{a}ndig}}
} \\
\parbox{33pt}{\raggedright 
\textsf{{\small 42 min}}
} & \parbox{400pt}{\raggedright 
\textsf{{\small L hilft Markus mit der Bedienung des Programms}}
} \\
\parbox{33pt}{\raggedright 
\textsf{{\small 44 min}}
} & \parbox{400pt}{\raggedright 
\textsf{{\small L geht weiter}}
} \\
\parbox{33pt}{\raggedright 
\textsf{{\small 45 min}}
} & \parbox{400pt}{\raggedright 
\textsf{{\small L hilft Alina mit der Simulation}}
} \\
\end{tabular}
\vspace{2pt}

}

\begin{indentation}{0pt}{0pt}{0pt}

\end{indentation}

\pagebreak

\section{Auswertung Quiz}

{\raggedright
\begin{indentation}{0pt}{0pt}{0pt}
Im folgenden werden zu jeder einzelnen Aufgabe die erzielten Punkte
aufgelistet bzw. nach "`Richtig"' und "`Falsch"' eingeteilt.
\end{indentation}
}

{\raggedright
\begin{indentation}{0pt}{0pt}{0pt}

\end{indentation}
}

{\raggedright

\vspace{3pt} \noindent
\begin{tabular}{p{74pt}p{31pt}p{4pt}p{35pt}p{8pt}p{49pt}p{-1pt}p{67pt}p{-1pt}}
\parbox{74pt}{\raggedright 
\textbf{\textsf{{\small Name}}}
} & \parbox{31pt}{\raggedright 
\textbf{\textsf{{\small Quiz 1}}}
} & \parbox{4pt}{\raggedright 
{\small \begin{math}\sum{}\end{math}}
} & \parbox{35pt}{\raggedright 
\textbf{\textsf{{\small Quiz 2}}}
} & \parbox{8pt}{\raggedright 
{\small \begin{math}\sum{}\end{math}}
} & \parbox{49pt}{\raggedright 
\textbf{\textsf{{\small Quiz 3}}}
} & \parbox{-1pt}{\raggedright 
{\small \begin{math}\sum{}\end{math}}
} & \parbox{67pt}{\raggedright 
\textbf{\textsf{{\small Quiz 4}}}
} & \parbox{-1pt}{\raggedright 
{\small \begin{math}\sum{}\end{math}}
} \\
\parbox{74pt}{\raggedright 
\textsf{{\small Ahmad}}
} & \parbox{31pt}{\raggedright 
\textsf{{\small FFFFF}}
} & \parbox{4pt}{\raggedright 
\textsf{{\small 0}}
} & \parbox{35pt}{\raggedright 
\textsf{{\small FFFFF}}
} & \parbox{8pt}{\raggedright 
\textsf{{\small 0}}
} & \parbox{49pt}{\raggedright 
\textsf{{\small FFFFFFFF}}
} & \parbox{-1pt}{\raggedright 
\textsf{{\small 0}}
} & \parbox{67pt}{\raggedright 
\textsf{{\small RFFFFFFFFF}}
} & \parbox{-1pt}{\raggedright 
\textsf{{\small 1}}
} \\
\parbox{74pt}{\raggedright 
\textsf{{\small Alina}}
} & \parbox{31pt}{\raggedright 
\textsf{{\small RRRFR}}
} & \parbox{4pt}{\raggedright 
\textsf{{\small 4}}
} & \parbox{35pt}{\raggedright 
\textsf{{\small RRFFR}}
} & \parbox{8pt}{\raggedright 
\textsf{{\small 3}}
} & \parbox{49pt}{\raggedright 
\textsf{{\small RFRFRFRF}}
} & \parbox{-1pt}{\raggedright 
\textsf{{\small 4}}
} & \parbox{67pt}{\raggedright 
\textsf{{\small RFRFRFRFRF}}
} & \parbox{-1pt}{\raggedright 
\textsf{{\small 5}}
} \\
\parbox{74pt}{\raggedright 
\textsf{{\small Artur}}
} & \parbox{31pt}{\raggedright 
\textsf{{\small FFFFF}}
} & \parbox{4pt}{\raggedright 
\textsf{{\small 0}}
} & \parbox{35pt}{\raggedright 
\textsf{{\small RRRRR}}
} & \parbox{8pt}{\raggedright 
\textsf{{\small 5}}
} & \parbox{49pt}{\raggedright 
\textsf{{\small FFFFFFFF}}
} & \parbox{-1pt}{\raggedright 
\textsf{{\small 0}}
} & \parbox{67pt}{\raggedright 
\textsf{{\small FFFFFFFFFF}}
} & \parbox{-1pt}{\raggedright 
\textsf{{\small 0}}
} \\
\parbox{74pt}{\raggedright 
\textsf{{\small Asgard}}
} & \parbox{31pt}{\raggedright 
\textsf{{\small RRRFR}}
} & \parbox{4pt}{\raggedright 
\textsf{{\small 4}}
} & \parbox{35pt}{\raggedright 
\textsf{{\small RRRFF}}
} & \parbox{8pt}{\raggedright 
\textsf{{\small 3}}
} & \parbox{49pt}{\raggedright 
\textsf{{\small RFRFRFFF}}
} & \parbox{-1pt}{\raggedright 
\textsf{{\small 3}}
} & \parbox{67pt}{\raggedright 
\textsf{{\small FFFFFFFFFF}}
} & \parbox{-1pt}{\raggedright 
\textsf{{\small 0}}
} \\
\parbox{74pt}{\raggedright 
\textsf{{\small Daniel}}
} & \parbox{31pt}{\raggedright 
\textsf{{\small RRRFR}}
} & \parbox{4pt}{\raggedright 
\textsf{{\small 4}}
} & \parbox{35pt}{\raggedright 
\textsf{{\small RRRRR}}
} & \parbox{8pt}{\raggedright 
\textsf{{\small 5}}
} & \parbox{49pt}{\raggedright 
\textsf{{\small FFRFFFFF}}
} & \parbox{-1pt}{\raggedright 
\textsf{{\small 1}}
} & \parbox{67pt}{\raggedright 
\textsf{{\small RRFRRFRRFR}}
} & \parbox{-1pt}{\raggedright 
\textsf{{\small 7}}
} \\
\parbox{74pt}{\raggedright 
\textsf{{\small Franziska}}
} & \parbox{31pt}{\raggedright 
\textsf{{\small RRRRR}}
} & \parbox{4pt}{\raggedright 
\textsf{{\small 5}}
} & \parbox{35pt}{\raggedright 
\textsf{{\small RRFFR}}
} & \parbox{8pt}{\raggedright 
\textsf{{\small 3}}
} & \parbox{49pt}{\raggedright 
\textsf{{\small RFRFRFRF}}
} & \parbox{-1pt}{\raggedright 
\textsf{{\small 4}}
} & \parbox{67pt}{\raggedright 
\textsf{{\small RRRRRFRFRF}}
} & \parbox{-1pt}{\raggedright 
\textsf{{\small 7}}
} \\
\parbox{74pt}{\raggedright 
\textsf{{\small Friederike}}
} & \parbox{31pt}{\raggedright 
\textsf{{\small RRRRF}}
} & \parbox{4pt}{\raggedright 
\textsf{{\small 4}}
} & \parbox{35pt}{\raggedright 
\textsf{{\small RRFFR}}
} & \parbox{8pt}{\raggedright 
\textsf{{\small 3}}
} & \parbox{49pt}{\raggedright 
\textsf{{\small RFRFRFFF}}
} & \parbox{-1pt}{\raggedright 
\textsf{{\small 3}}
} & \parbox{67pt}{\raggedright 
\textsf{{\small RFRFFFRFFF}}
} & \parbox{-1pt}{\raggedright 
\textsf{{\small 3}}
} \\
\parbox{74pt}{\raggedright 
\textsf{{\small Janna}}
} & \parbox{31pt}{\raggedright 

} & \parbox{4pt}{\raggedright 

} & \parbox{35pt}{\raggedright 
\textsf{{\small RRRRR}}
} & \parbox{8pt}{\raggedright 
\textsf{{\small 5}}
} & \parbox{49pt}{\raggedright 
\textsf{{\small FFRFFFFF}}
} & \parbox{-1pt}{\raggedright 
\textsf{{\small 1}}
} & \parbox{67pt}{\raggedright 
\textsf{{\small FFFFFFFFFF}}
} & \parbox{-1pt}{\raggedright 
\textsf{{\small 0}}
} \\
\parbox{74pt}{\raggedright 
\textsf{{\small Markus}}
} & \parbox{31pt}{\raggedright 
\textsf{{\small RRFFF}}
} & \parbox{4pt}{\raggedright 
\textsf{{\small 2}}
} & \parbox{35pt}{\raggedright 
\textsf{{\small RRRFF}}
} & \parbox{8pt}{\raggedright 
\textsf{{\small 3}}
} & \parbox{49pt}{\raggedright 
\textsf{{\small FFFFFFFF}}
} & \parbox{-1pt}{\raggedright 
\textsf{{\small 0}}
} & \parbox{67pt}{\raggedright 
\textsf{{\small FFFFFFFFFF}}
} & \parbox{-1pt}{\raggedright 
\textsf{{\small 0}}
} \\
\parbox{74pt}{\raggedright 
\textsf{{\small Pascal}}
} & \parbox{31pt}{\raggedright 

} & \parbox{4pt}{\raggedright 

} & \parbox{35pt}{\raggedright 
\textsf{{\small RFFFF}}
} & \parbox{8pt}{\raggedright 
\textsf{{\small 1}}
} & \parbox{49pt}{\raggedright 
\textsf{{\small FFFFFFFF}}
} & \parbox{-1pt}{\raggedright 
\textsf{{\small 0}}
} & \parbox{67pt}{\raggedright 
\textsf{{\small FFFFFFFFFF}}
} & \parbox{-1pt}{\raggedright 
\textsf{{\small 0}}
} \\
\parbox{74pt}{\raggedright 
\textsf{{\small Per-S\"{o}ren}}
} & \parbox{31pt}{\raggedright 
\textsf{{\small FFFFF}}
} & \parbox{4pt}{\raggedright 
\textsf{{\small 0}}
} & \parbox{35pt}{\raggedright 

} & \parbox{8pt}{\raggedright 

} & \parbox{49pt}{\raggedright 

} & \parbox{-1pt}{\raggedright 

} & \parbox{67pt}{\raggedright 
\textsf{{\small RFFFFFFFFF}}
} & \parbox{-1pt}{\raggedright 
\textsf{{\small 1}}
} \\
\parbox{74pt}{\raggedright 
\textsf{{\small Quoc-Thai}}
} & \parbox{31pt}{\raggedright 
\textsf{{\small FFFFF}}
} & \parbox{4pt}{\raggedright 
\textsf{{\small 0}}
} & \parbox{35pt}{\raggedright 
\textsf{{\small RRRRR}}
} & \parbox{8pt}{\raggedright 
\textsf{{\small 5}}
} & \parbox{49pt}{\raggedright 
\textsf{{\small FFFFFFFF}}
} & \parbox{-1pt}{\raggedright 
\textsf{{\small 0}}
} & \parbox{67pt}{\raggedright 
\textsf{{\small FFFFFFFFFF}}
} & \parbox{-1pt}{\raggedright 
\textsf{{\small 0}}
} \\
\parbox{74pt}{\raggedright 
\textsf{{\small Steven}}
} & \parbox{31pt}{\raggedright 
\textsf{{\small RRRFR}}
} & \parbox{4pt}{\raggedright 
\textsf{{\small 4}}
} & \parbox{35pt}{\raggedright 
\textsf{{\small RRRFR}}
} & \parbox{8pt}{\raggedright 
\textsf{{\small 4}}
} & \parbox{49pt}{\raggedright 
\textsf{{\small RFRFFFFF}}
} & \parbox{-1pt}{\raggedright 
\textsf{{\small 2}}
} & \parbox{67pt}{\raggedright 
\textsf{{\small RRRRRFRRFF}}
} & \parbox{-1pt}{\raggedright 
\textsf{{\small 7}}
} \\
\end{tabular}
\vspace{2pt}

}

\begin{indentation}{0pt}{0pt}{0pt}

\end{indentation}

{\raggedright

\vspace{3pt} \noindent
\begin{tabular}{p{74pt}p{31pt}p{4pt}p{35pt}p{13pt}p{44pt}p{4pt}p{58pt}}
\parbox{74pt}{\raggedright 
\textbf{\textsf{{\small Name}}}
} & \parbox{31pt}{\raggedright 
\textbf{\textsf{{\small Quiz 5}}}
} & \parbox{4pt}{\raggedright 
\textbf{{\small \begin{math}\sum{}\end{math}}}
} & \parbox{35pt}{\raggedright 
\textbf{\textsf{{\small Quiz 6}}}
} & \parbox{13pt}{\raggedright 
\textbf{{\small \begin{math}\sum{}\end{math}}}
} & \parbox{44pt}{\raggedright 
\textbf{\textsf{{\small Quiz 7}}}
} & \parbox{4pt}{\raggedright 
\textbf{{\small \begin{math}\sum{}\end{math}}}
} & \parbox{58pt}{\raggedright 
\textbf{\textsf{{\small Final }}}

\textbf{\textsf{{\small Exam}}}
} \\
\parbox{74pt}{\raggedright 
\textsf{{\small Ahmad}}
} & \parbox{31pt}{\raggedright 

} & \parbox{4pt}{\raggedright 

} & \parbox{35pt}{\raggedright 
\textsf{{\small 10100}}
} & \parbox{13pt}{\raggedright 
\textsf{{\small 2}}
} & \parbox{44pt}{\raggedright 
\textsf{{\small 10001}}
} & \parbox{4pt}{\raggedright 
\textsf{{\small 2}}
} & \parbox{58pt}{\raggedright 
\textsf{{\small 211133}}
} \\
\parbox{74pt}{\raggedright 
\textsf{{\small Alina}}
} & \parbox{31pt}{\raggedright 
\textsf{{\small 111}}
} & \parbox{4pt}{\raggedright 
\textsf{{\small 3}}
} & \parbox{35pt}{\raggedright 
\textsf{{\small 11114}}
} & \parbox{13pt}{\raggedright 
\textsf{{\small 8}}
} & \parbox{44pt}{\raggedright 
\textsf{{\small 11114}}
} & \parbox{4pt}{\raggedright 
\textsf{{\small 8}}
} & \parbox{58pt}{\raggedright 
\textsf{{\small 211133}}
} \\
\parbox{74pt}{\raggedright 
\textsf{{\small Artur}}
} & \parbox{31pt}{\raggedright 
\textsf{{\small 121}}
} & \parbox{4pt}{\raggedright 
\textsf{{\small 4}}
} & \parbox{35pt}{\raggedright 
\textsf{{\small 01010}}
} & \parbox{13pt}{\raggedright 
\textsf{{\small 2}}
} & \parbox{44pt}{\raggedright 
\textsf{{\small 11004}}
} & \parbox{4pt}{\raggedright 
\textsf{{\small 6}}
} & \parbox{58pt}{\raggedright 
\textsf{{\small 201133}}
} \\
\parbox{74pt}{\raggedright 
\textsf{{\small Asgard}}
} & \parbox{31pt}{\raggedright 

} & \parbox{4pt}{\raggedright 

} & \parbox{35pt}{\raggedright 
\textsf{{\small 00011}}
} & \parbox{13pt}{\raggedright 
\textsf{{\small 2}}
} & \parbox{44pt}{\raggedright 

} & \parbox{4pt}{\raggedright 

} & \parbox{58pt}{\raggedright 
\textsf{{\small 101103}}
} \\
\parbox{74pt}{\raggedright 
\textsf{{\small Daniel}}
} & \parbox{31pt}{\raggedright 
\textsf{{\small 120}}
} & \parbox{4pt}{\raggedright 
\textsf{{\small 3}}
} & \parbox{35pt}{\raggedright 
\textsf{{\small 01102}}
} & \parbox{13pt}{\raggedright 
\textsf{{\small 4}}
} & \parbox{44pt}{\raggedright 
\textsf{{\small 11013}}
} & \parbox{4pt}{\raggedright 
\textsf{{\small 6}}
} & \parbox{58pt}{\raggedright 
\textsf{{\small 211133}}
} \\
\parbox{74pt}{\raggedright 
\textsf{{\small Franziska}}
} & \parbox{31pt}{\raggedright 
\textsf{{\small 100}}
} & \parbox{4pt}{\raggedright 
\textsf{{\small 1}}
} & \parbox{35pt}{\raggedright 
\textsf{{\small 11114}}
} & \parbox{13pt}{\raggedright 
\textsf{{\small 8}}
} & \parbox{44pt}{\raggedright 
\textsf{{\small 11114}}
} & \parbox{4pt}{\raggedright 
\textsf{{\small 8}}
} & \parbox{58pt}{\raggedright 
\textsf{{\small 211123}}
} \\
\parbox{74pt}{\raggedright 
\textsf{{\small Friederike}}
} & \parbox{31pt}{\raggedright 
\textsf{{\small 100}}
} & \parbox{4pt}{\raggedright 
\textsf{{\small 1}}
} & \parbox{35pt}{\raggedright 
\textsf{{\small 11110}}
} & \parbox{13pt}{\raggedright 
\textsf{{\small 4}}
} & \parbox{44pt}{\raggedright 
\textsf{{\small 11113}}
} & \parbox{4pt}{\raggedright 
\textsf{{\small 7}}
} & \parbox{58pt}{\raggedright 
\textsf{{\small 211133}}
} \\
\parbox{74pt}{\raggedright 
\textsf{{\small Janna}}
} & \parbox{31pt}{\raggedright 
\textsf{{\small 010}}
} & \parbox{4pt}{\raggedright 
\textsf{{\small 1}}
} & \parbox{35pt}{\raggedright 
\textsf{{\small 00002}}
} & \parbox{13pt}{\raggedright 
\textsf{{\small 2}}
} & \parbox{44pt}{\raggedright 
\textsf{{\small 11002}}
} & \parbox{4pt}{\raggedright 
\textsf{{\small 4}}
} & \parbox{58pt}{\raggedright 
\textsf{{\small 111022}}
} \\
\parbox{74pt}{\raggedright 
\textsf{{\small Markus}}
} & \parbox{31pt}{\raggedright 
\textsf{{\small 100}}
} & \parbox{4pt}{\raggedright 
\textsf{{\small 1}}
} & \parbox{35pt}{\raggedright 
\textsf{{\small 00110}}
} & \parbox{13pt}{\raggedright 
\textsf{{\small 2}}
} & \parbox{44pt}{\raggedright 
\textsf{{\small 00111}}
} & \parbox{4pt}{\raggedright 
\textsf{{\small 3}}
} & \parbox{58pt}{\raggedright 
\textsf{{\small 111103}}
} \\
\parbox{74pt}{\raggedright 
\textsf{{\small Pascal}}
} & \parbox{31pt}{\raggedright 
\textsf{{\small 200}}
} & \parbox{4pt}{\raggedright 
\textsf{{\small 2}}
} & \parbox{35pt}{\raggedright 
\textsf{{\small 01110}}
} & \parbox{13pt}{\raggedright 
\textsf{{\small 3}}
} & \parbox{44pt}{\raggedright 
\textsf{{\small 00000}}
} & \parbox{4pt}{\raggedright 
\textsf{{\small 0}}
} & \parbox{58pt}{\raggedright 
\textsf{{\small 201133}}
} \\
\parbox{74pt}{\raggedright 
\textsf{{\small Per-S\"{o}ren}}
} & \parbox{31pt}{\raggedright 
\textsf{{\small 000}}
} & \parbox{4pt}{\raggedright 
\textsf{{\small 0}}
} & \parbox{35pt}{\raggedright 
\textsf{{\small 10100}}
} & \parbox{13pt}{\raggedright 
\textsf{{\small 2}}
} & \parbox{44pt}{\raggedright 
\textsf{{\small 10101}}
} & \parbox{4pt}{\raggedright 
\textsf{{\small 3}}
} & \parbox{58pt}{\raggedright 
\textsf{{\small 211133}}
} \\
\parbox{74pt}{\raggedright 
\textsf{{\small Quoc-Thai}}
} & \parbox{31pt}{\raggedright 
\textsf{{\small 100}}
} & \parbox{4pt}{\raggedright 
\textsf{{\small 1}}
} & \parbox{35pt}{\raggedright 
\textsf{{\small 11100}}
} & \parbox{13pt}{\raggedright 
\textsf{{\small 3}}
} & \parbox{44pt}{\raggedright 
\textsf{{\small 00002}}
} & \parbox{4pt}{\raggedright 
\textsf{{\small 2}}
} & \parbox{58pt}{\raggedright 
\textsf{{\small 211133}}
} \\
\parbox{74pt}{\raggedright 
\textsf{{\small Steven}}
} & \parbox{31pt}{\raggedright 
\textsf{{\small 211}}
} & \parbox{4pt}{\raggedright 
\textsf{{\small 4}}
} & \parbox{35pt}{\raggedright 
\textsf{{\small 00001}}
} & \parbox{13pt}{\raggedright 
\textsf{{\small 1}}
} & \parbox{44pt}{\raggedright 
\textsf{{\small 11014}}
} & \parbox{4pt}{\raggedright 
\textsf{{\small 7}}
} & \parbox{58pt}{\raggedright 
\textsf{{\small 211133}}
} \\
\end{tabular}
\vspace{2pt}

}

\begin{indentation}{0pt}{0pt}{0pt}

\end{indentation}

\pagebreak{}

\section{Arbeitsbl\"{a}tter}

{\raggedleft
\begin{indentation}{0pt}{0pt}{0pt}
{\scriptsize Dr. Schmekel (Cornell U.)}
\end{indentation}
}

{\raggedleft
\begin{indentation}{0pt}{0pt}{0pt}
{\scriptsize Physics GK S2 \textendash{} Spring 2009}
\end{indentation}
}

{\raggedleft
\begin{indentation}{0pt}{0pt}{0pt}

\end{indentation}
}

{\raggedleft
\begin{indentation}{0pt}{0pt}{0pt}

\end{indentation}
}

\begin{center}
\begin{indentation}{0pt}{0pt}{0pt}
\textbf{{\large Instructions for using PSPICE}}
\end{indentation}
\end{center}

{\raggedright
\begin{indentation}{0pt}{0pt}{0pt}

\end{indentation}
}

{\raggedright
\begin{indentation}{0pt}{0pt}{0pt}

\end{indentation}
}

\begin{enumerate}
	\item Copy the file ``test.cir'' from the
provided USB-stick onto your laptop.
\end{enumerate}

{\raggedright
\begin{indentation}{36pt}{0pt}{0pt}

\end{indentation}
}

\begin{enumerate}
	\item Start the application ``PSPICE AD''.
\end{enumerate}

{\raggedright
\begin{indentation}{36pt}{0pt}{0pt}

\end{indentation}
}

\begin{enumerate}
	\item Open the test file. You have to choose *.cir \textendash{} files or
all files *.* in order to see the file in your chosen directory.
\end{enumerate}

{\raggedright
\begin{indentation}{36pt}{0pt}{0pt}

\end{indentation}
}

\begin{enumerate}
	\item Click on the item ``Run'' in the menu item ``Simulation''.
\end{enumerate}

{\raggedright
\begin{indentation}{36pt}{0pt}{0pt}

\end{indentation}
}

\begin{enumerate}
	\item View the output file by clicking on ``Output file'' in the menu item ``View''.
\end{enumerate}

{\raggedright
\begin{indentation}{36pt}{0pt}{0pt}

\end{indentation}
}

\begin{enumerate}
	\item Draw a circuit diagram of the circuit you would like to simulate. The
circuit described by the test file looks like this
\end{enumerate}

{\raggedright
\begin{indentation}{0pt}{0pt}{0pt}

\end{indentation}
}

{\raggedright
\begin{indentation}{0pt}{0pt}{0pt}

\end{indentation}
}

{\raggedright
\begin{indentation}{0pt}{0pt}{0pt}

\end{indentation}
}

{\raggedright
\begin{indentation}{0pt}{0pt}{0pt}

\end{indentation}
}

{\raggedright
\begin{indentation}{70pt}{0pt}{0pt}
\includegraphics[width=256pt]{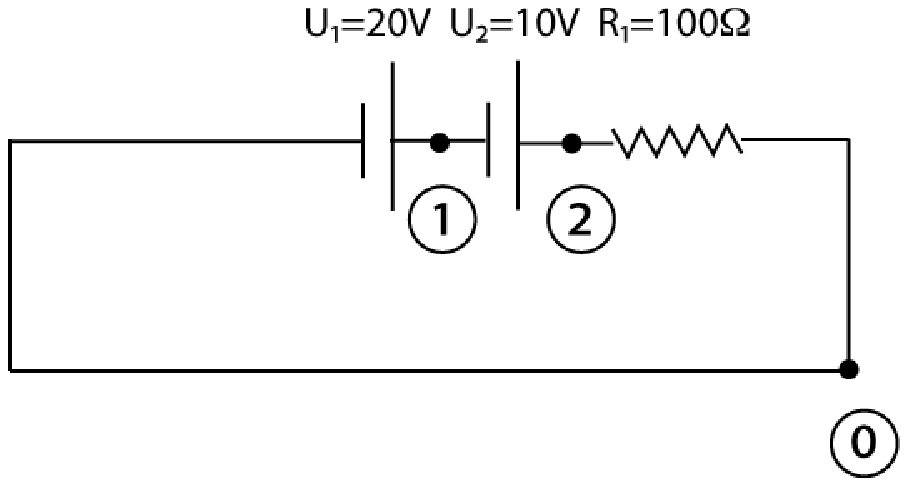}

\end{indentation}
}

{\raggedright
\begin{indentation}{0pt}{0pt}{0pt}

\end{indentation}
}

{\raggedright
\begin{indentation}{0pt}{0pt}{0pt}

\end{indentation}
}

\begin{enumerate}
	\item Label all the nodes, modify the test file accordingly and run the
simulation.
\end{enumerate}

{\raggedright
\begin{indentation}{36pt}{0pt}{0pt}

\end{indentation}
}

{\raggedright
\begin{indentation}{0pt}{0pt}{0pt}
\pagebreak{}

\end{indentation}
}

{\raggedleft
\begin{indentation}{0pt}{0pt}{0pt}
{\scriptsize Dr. Schmekel (Cornell U.)}
\end{indentation}
}

{\raggedleft
\begin{indentation}{0pt}{0pt}{0pt}
{\scriptsize Physics GK S2 \textendash{} Spring 2009}
\end{indentation}
}

{\raggedleft
\begin{indentation}{0pt}{0pt}{0pt}

\end{indentation}
}

{\raggedleft
\begin{indentation}{0pt}{0pt}{0pt}

\end{indentation}
}

\begin{center}
\begin{indentation}{0pt}{0pt}{0pt}
\textbf{{\large Time-domain analysis with PSPICE (``Transient Analysis'')}}
\end{indentation}
\end{center}

{\raggedright
\begin{indentation}{0pt}{0pt}{0pt}

\end{indentation}
}

{\raggedright
\begin{indentation}{0pt}{0pt}{0pt}

\end{indentation}
}

\begin{enumerate}
	\item Copy the file ``cap.cir'' from the
provided USB-stick onto your laptop.
\end{enumerate}

{\raggedright
\begin{indentation}{36pt}{0pt}{0pt}

\end{indentation}
}

{\raggedright
\begin{indentation}{36pt}{0pt}{0pt}
This is the contents of the file
\end{indentation}
}

{\raggedright
\begin{indentation}{36pt}{0pt}{0pt}

\end{indentation}
}

{\raggedright
\begin{indentation}{36pt}{0pt}{0pt}
.WIDTH OUT=80
\end{indentation}
}

{\raggedright
\begin{indentation}{36pt}{0pt}{0pt}
V1 1 0 PULSE 0 10 1 0 0 100 100
\end{indentation}
}

{\raggedright
\begin{indentation}{36pt}{0pt}{0pt}
R1 1 2 1k
\end{indentation}
}

{\raggedright
\begin{indentation}{36pt}{0pt}{0pt}
C1 2 0 100u
\end{indentation}
}

{\raggedright
\begin{indentation}{36pt}{0pt}{0pt}
.TRAN 0.01us 3s
\end{indentation}
}

{\raggedright
\begin{indentation}{36pt}{0pt}{0pt}
.PROBE
\end{indentation}
}

{\raggedright
\begin{indentation}{36pt}{0pt}{0pt}
.END
\end{indentation}
}

{\raggedright
\begin{indentation}{36pt}{0pt}{0pt}

\end{indentation}
}

{\raggedright
\begin{indentation}{36pt}{0pt}{0pt}
and this is the circuit described by the file
\end{indentation}
}

{\raggedright
\begin{indentation}{36pt}{0pt}{0pt}

\end{indentation}
}

\begin{center}
\begin{indentation}{36pt}{0pt}{0pt}
\includegraphics[width=256pt]{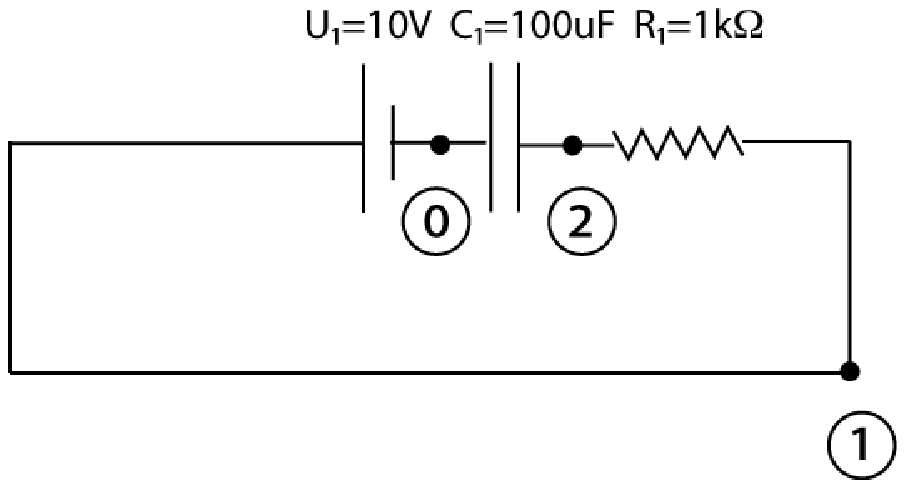}

\end{indentation}
\end{center}

{\raggedright
\begin{indentation}{36pt}{0pt}{0pt}

\end{indentation}
}

\begin{enumerate}
	\item Start the application ``PSPICE AD''.
\end{enumerate}

{\raggedright
\begin{indentation}{36pt}{0pt}{0pt}

\end{indentation}
}

\begin{enumerate}
	\item Open the test file. You have to choose *.cir \textendash{} files or
all files *.* in order to see the file in your chosen directory.
\end{enumerate}

{\raggedright
\begin{indentation}{36pt}{0pt}{0pt}

\end{indentation}
}

\begin{enumerate}
	\item Click on the item ``Run'' in the menu item ``Simulation''.
\end{enumerate}

{\raggedright
\begin{indentation}{36pt}{0pt}{0pt}

\end{indentation}
}

\begin{enumerate}
	\item Choose ``Add trace'' from the menu
item ``Trace'' and select the quantity you are interested in.
\end{enumerate}

{\raggedright
\begin{indentation}{36pt}{0pt}{0pt}

\end{indentation}
}

{\raggedright
\begin{indentation}{0pt}{0pt}{0pt}

\end{indentation}
}

{\raggedright
\begin{indentation}{0pt}{0pt}{0pt}

\end{indentation}
}

{\raggedright
\begin{indentation}{0pt}{0pt}{0pt}

\end{indentation}
}

{\raggedright
\begin{indentation}{0pt}{0pt}{0pt}

\end{indentation}
}

{\raggedright
\begin{indentation}{0pt}{0pt}{0pt}

\end{indentation}
}

{\raggedright
\begin{indentation}{0pt}{0pt}{0pt}

\end{indentation}
}

\pagebreak{}

\section{PSPICE - Beispiel f\"{u}r Ein- und Ausgabe}

{\raggedright
\begin{indentation}{0pt}{0pt}{0pt}

\end{indentation}
}

\subsection{Eingabedatei}

{\raggedright
\begin{indentation}{0pt}{0pt}{0pt}

\end{indentation}
}

{\raggedright
\begin{indentation}{0pt}{0pt}{0pt}
* Testschaltung
\end{indentation}
}

{\raggedright
\begin{indentation}{0pt}{0pt}{0pt}
.WIDTH OUT=80
\end{indentation}
}

{\raggedright
\begin{indentation}{0pt}{0pt}{0pt}
V1 1 0 PULSE 0 10 1 0 0 100 100
\end{indentation}
}

{\raggedright
\begin{indentation}{0pt}{0pt}{0pt}
R1 1 2 1k
\end{indentation}
}

{\raggedright
\begin{indentation}{0pt}{0pt}{0pt}
C1 2 0 100u
\end{indentation}
}

{\raggedright
\begin{indentation}{0pt}{0pt}{0pt}
.TRAN 0.01us 3s
\end{indentation}
}

{\raggedright
\begin{indentation}{0pt}{0pt}{0pt}
.PROBE
\end{indentation}
}

{\raggedright
\begin{indentation}{0pt}{0pt}{0pt}
.END
\end{indentation}
}

{\raggedright
\begin{indentation}{0pt}{0pt}{0pt}

\end{indentation}
}

\subsection{Ausgabe}

\begin{indentation}{0pt}{0pt}{0pt}
\includegraphics[width=467pt,angle=270]{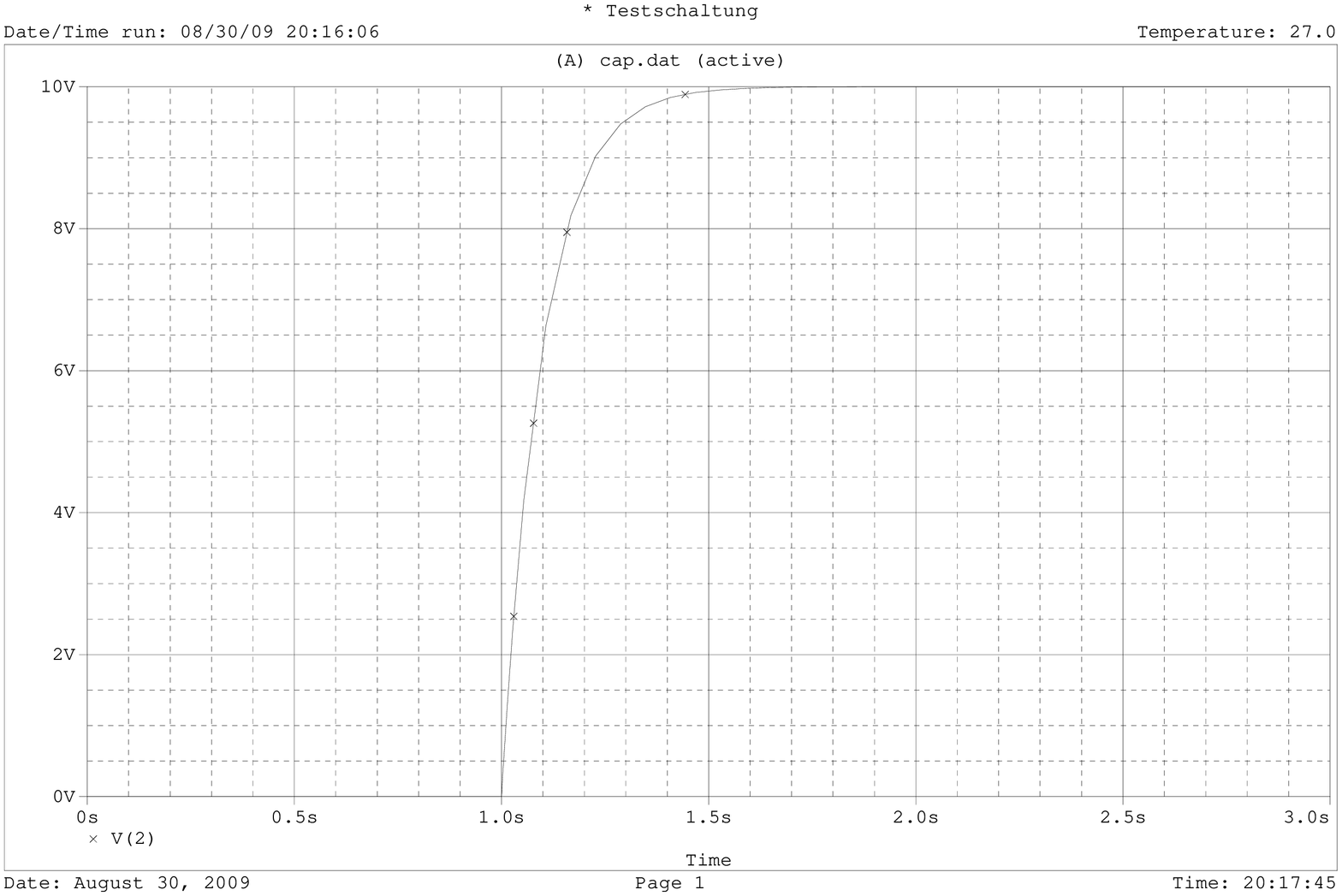}

\end{indentation}

\begin{indentation}{0pt}{0pt}{0pt}
{\small Der Graph zeigt die Spannung am Knoten 2 (bez\"{u}glich Knoten
0) f\"{u}r die Schaltung beschrieben durch die Datei cap.cir aus Anhang
D.}
\end{indentation}

\end{document}